\documentstyle[11pt,aaspp4]{article}

 \tighten

\slugcomment{preprint astro-ph/9512122; ApJ, in press}

\lefthead{Sackmann \& Boothroyd}
\righthead{${}^3$He, ${}^7$Li, Be, and B in Red Giants}

     \def\bootplotfidtwo#1#2#3#4#5#6#7#8#9{\centering \leavevmode
       \vbox to#3{\rule{0pt}{#3}}
       \includegraphics{#1}
       \hfil
       \includegraphics{#2}}

\begin{document}

\title{Creation of ${}^7$Li and Destruction of ${}^3$He, ${}^9$Be,
 ${}^{10}$B, and ${}^{11}$B in Low Mass Red Giants, Due to Deep Circulation}

\author{I.-Juliana Sackmann}
\affil{W. K. Kellogg Radiation Laboratory 106-38,
 California Institute of Technology,\\
 Pasadena, CA 91125}
\authoremail{ijs@krl.caltech.edu}

\and

\author{Arnold I. Boothroyd\altaffilmark{1}}
\authoremail{aib@krl.caltech.edu}
\affil{Dept.\ of Mathematics \& Statistics, Monash University,
 Clayton, VIC~3168, Australia}
\altaffiltext{1}{now at W.~K.~Kellogg Radiation Laboratory \hbox{106-38},
 California Institute of Technology, Pasadena, CA 91125}

\begin{abstract}

It has been demonstrated that \hbox{$\rm{}^7{Li}$}
{\it can be created in low mass red giant stars} via the
Cameron-Fowler mechanism, due to
extra deep mixing and the associated ``{\it cool bottom processing\/}''.
Under certain conditions, this \hbox{$\rm{}^7{Li}$} creation can take the
place of the \hbox{$\rm{}^7{Li}$} destruction normally expected.
Note that such extra mixing on the red giant branch (RGB) has previously
been invoked to explain the observed \hbox{$\rm{}^{13}C$} enhancement.
This new \hbox{$\rm{}^7{Li}$} production can account for
the recent discovery of surprisingly high lithium
abundances in some low mass red giants (a few of which are super-rich
lithium stars, with abundances higher than that in the interstellar
medium).  The amount of \hbox{$\rm{}^7{Li}$} produced can exceed
$\log\,\varepsilon({}^7{\rm Li}) \sim 4$, but
depends critically on the details of
the extra mixing mechanism (mixing speeds, geometry, episodicity).
If the deep circulation is a relatively long-lived, continuous process,
lithium-rich RGB stars should be completely devoid of beryllium and boron.

Cool bottom processing leads to \hbox{$\rm{}^3{He}$} destruction in low
mass stars; in contrast to the \hbox{$\rm{}^7{Li}$} creation, the
extent of \hbox{$\rm{}^3{He}$} depletion is largely
independent of the details of the extra mixing mechanism.
The overall contribution from solar-metallicity stars
(from 1 to $40\>M_\odot$)
is expected to be a net destruction of~\hbox{$\rm{}^3{He}$}, with an
overall \hbox{$\rm{}^3{He}$} survival fraction $g_3 \approx 0.9 \pm 0.2$
(weighted average over all stellar masses); this is in contrast to
the conclusion from standard dredge-up theory, which would predict that
stars are net producers of~\hbox{$\rm{}^3{He}$}
(with~$g_3^{\rm dr} \sim 2.4 \pm 0.5$).
Population~II stars experience even more severe \hbox{$\rm{}^3{He}$}
depletion, with $0.3 \lesssim g_3 \lesssim 0.7$.
Destruction of \hbox{$\rm{}^3{He}$} in low mass stars is consistent with
the requirements of galactic chemical evolution models; it would also
result in some relaxation of the upper bound on the primordial
(D+\hbox{$\rm{}^3{He}$})/H abundance, thus relaxing the lower bound
on the cosmic baryon density~$\Omega_b$
from Big Bang nucleosynthesis calculations.

For reference, we also present the effects of standard first and second
dredge-up on the helium, lithium, beryllium, and boron isotopes.
\end{abstract}



\section{Introduction} \label{intro}

For some time, it has been known that stars super-rich in lithium exist,
with abundances much higher than they could have been endowed with at
birth, i.e, larger than the present interstellar medium value.
A much clearer understanding of these super-rich lithium stars
was provided by the recent Magellanic Cloud observations of
Smith \& Lambert (1989\markcite{SmiL89}, 1990\markcite{SmiL90}),
who obtained for the first time a luminosity range
for lithium-rich asymptotic giant branch (AGB) stars.  They found that
the most luminous AGB stars were lithium-rich, in excellent agreement with
the predictions of theoretical models of hot bottom burning
(Sackmann, Smith, \& Despain 1974\markcite{SSD74};
Scalo, Despain, \& Ulrich 1975\markcite{ScaDU75};
Sackmann \& Boothroyd 1992\markcite{SB92}, 1995\markcite{SB95},
1998\markcite{SB98}).
Hot bottom burning is predicted to occur
occur only in the most massive AGB stars (between $\sim 4$ and
$\sim 7\>M_\odot$), which are also the most luminous ones
(Iben 1975\markcite{Ib75};
Bl\"ocker \& Sch\"onberner 1991\markcite{BloS91};
Lattanzio 1992\markcite{Lat92};
Sackmann \& Boothroyd 1992\markcite{SB92}, 1995\markcite{SB95},
1998\markcite{SB98};
Boothroyd \& Sackmann 1992\markcite{BS92};
Boothroyd, Sackmann, \& Ahern 1993\markcite{BSA93}).
However, some lithium-rich stars do not fit into the above scenario.

During most of the evolution of a star, lithium is destroyed, as it
burns at relatively low temperatures.  During the pre-main sequence phase,
there can be significant lithium destruction for stars of
masses $\lesssim 1\>M_\odot$ (see, e.g.,
D'Antona \& Mazzitelli 1984\markcite{DAntM84};
Proffitt \& Michaud 1989\markcite{ProM89};
VandenBerg \& Poll 1989\markcite{VandP89};
Michaud \& Charbonneau 1991\markcite{MicC91}).
During the main sequence phase, considerable lithium destruction is
observed in low mass stars ($\lesssim 1.2\>M_\odot$), e.g., by two orders
of magnitude for the Sun; this is generally explained by rotation-induced
slow mixing, with some diffusion (see, e.g.,
Schatzman 1977\markcite{Scha77};
Baglin, Morel, \& Schatzman 1985\markcite{BagMS85};
Vauclair 1988\markcite{Vau88};
Pinsonneault et al.\ 1989\markcite{Pin+89};
Michaud \& Charbonneau 1991\markcite{MicC91};
Proffitt \& Michaud 1991\markcite{ProM91};
Charbonnel, Vauclair, \& Zahn 1992\markcite{CharVZ92};
Chaboyer, Demarque, \& Pinsonneault 1995\markcite{ChabDP95}),
though part of this effect might possibly be due to early main sequence
mass loss
(Boothroyd, Sackmann, \& Fowler 1991\markcite{BSF91};
Swenson \& Faulkner 1992\markcite{SweF92};
Whitmire et al.\ 1995\markcite{Whit+95};
Guzik \& Cox 1995\markcite{GuzC95}).
As stars approach the base of the red giant branch (RGB),
the surface lithium abundance declines further by two orders of magnitude,
as the deepening convective envelope reaches into lithium-depleted
interior layers.  Thus RGB stars, and AGB stars of $\lesssim 4\>M_\odot$,
are expected to have surface lithium abundances from two to four orders
of magnitude below the present interstellar medium value of
$\hbox{$\log\,\varepsilon({}^7{\rm Li})$} \sim 3.3$
(where $\hbox{$\log\,\varepsilon({}^7{\rm Li})$} \equiv
\log[n(\hbox{$\rm{}^7{Li}$})/n(\rm H)] + 12$).

In general, this is observed to be the case; Population~I red giants
usually have lithium abundances lying in the range $-1 \lesssim
\hbox{$\log\,\varepsilon({}^7{\rm Li})$} \lesssim 1$
(Lambert, Dominy, \& Sivertsen 1980\markcite{LamDS80};
Brown et al.\ 1989\markcite{Bro+89}).
However, a few (of order~1\%) of the red giants have been observed to
have lithium abundances excess of the standard predictions
(Brown et al.\ 1989\markcite{Bro+89}; see also
Wallerstein \& Sneden 1982\markcite{WalS82};
Hanni 1984\markcite{Han84};
Gratton \& D'Antona 1989\markcite{GraD89};
Pilachowski, Sneden, \& Hudek 1990\markcite{PilSH90};
Pallavicini et al.\ 1990\markcite{Pal+90};
Fekel \& Marschall 1991\markcite{FekM91};
Fekel \& Balachandran 1993\markcite{FekB93}),
occasionally having abundances much higher than the present interstellar
medium abundance
(da~Silva, de~la~Reza, \& Barbuy 1995a\markcite{daSRB95a},b\markcite{daSRB95b};
de~la~Reza \& da~Silva 1995\markcite{delaRS95};
de~la~Reza, Drake, \& da~Silva 1996\markcite{delaRDS96}).
Frequently, these lithium-rich red giants have large infrared excesses,
interpreted as associated circumstellar dust shells.  Recently,
de~la~Reza et al.\ (1997)\markcite{delaR+97}
discovered 20 new lithium-rich giants (doubling the total number known)
by searching for lithium lines in giants with infrared excesses.
These lithium-rich giants have
not yet reached the AGB, and at least some of them are observed to be
low mass stars; thus they cannot have experienced \hbox{$\rm{}^7{Li}$}
creation via hot bottom burning.
This presents a new puzzle, namely, how to explain such high lithium
abundances in pre-AGB stars.  A related puzzle is presented by observations
of field Population~II giants by
Pilachowski, Sneden, \& Booth (1993)\markcite{PilSB93};
they find extra lithium {\it depletion\/} by large factors occurring on the RGB
{\it subsequent\/} to first dredge-up.

Another problem concerns \hbox{$\rm{}^3{He}$}.  It is well known that
deuterium~(D) and \hbox{$\rm{}^3{He}$} are created in the Big Bang.  Stars
burn their initial~D to \hbox{$\rm{}^3{He}$} while still on
the pre-main sequence.  Low mass stars create \hbox{$\rm{}^3{He}$} pockets
in their interior during main sequence burning, as first pointed out by
Iben (1967)\markcite{Ib67},
which will subsequently be dredged up to the surface on the RGB, and
injected into the interstellar medium via stellar mass loss on the RGB
and AGB\hbox{}.  Recent measurements of \hbox{$\rm{}^3{He}$} mass
fractions of~$\sim 0.001$ in the low-mass planetary nebulae NGC~3242
and IC~289 confirm that this does indeed occur in at least some stars
(Galli et al.\ 1997\markcite{Gal+97}).
The sum of D+\hbox{$\rm{}^3{He}$} in the
interstellar medium was predicted to increase with time, due to this
stellar processing.  The observed solar ratio
$\rm (D{+}\hbox{$\rm{}^3{He}$})/H \approx 4 \times 10^{-5}$
has thus been used as an upper bound on the primordial ratio, in
order constrain Big Bang nucleosynthesis models (see, e.g.,
Steigman 1985\markcite{Ste85});
it provides the strongest
lower limit on the baryon to photon number ratio~$\eta \ge 3 \times
10^{-10}$, and thus to the baryon density of the universe~$\Omega_b \ge
0.01 h^{-2}$ (where $h$ is the Hubble constant in units of
$100\>$km$\,$s$^{-1}\,$Mpc$^{-1}$).

Rood, Bania, \& Wilson (1984)\markcite{RooBW84}
pointed out that the apparent observed trend with galactocentric radius
of the \hbox{$\rm{}^3{He}$} abundances in \ion{H}{2} regions
was the opposite of what one
would expect if the galactic \hbox{$\rm{}^3{He}$} abundance was increasing
with time due to stellar processing.  More recent data
(Rood et al.\ 1997\markcite{Roo+97};
Bania et al.\ 1997\markcite{Ban+97})
show no significant trend of the \hbox{$\rm{}^3{He}$} abundance in \ion{H}{2}
regions as a function of either galactocentric radius or metallicity, with
an average abundance of $\rm {}^3 He / H = 1.5^{+1.0}_{-0.5} \times 10^{-5}$.
Some extragalactic measurements of deuterium have recently appeared;
a high ratio
$\rm D/H \approx 2 \times 10^{-4}$
recently observed in a high-redshift ($z=3.32$) absorption system in the
quasar Q0014+813
(Songaila et al.\ 1994\markcite{Son+94};
Carswell et al.\ 1994\markcite{Cars+94};
Rugers \& Hogan 1996a\markcite{RugH96a})
has sparked new interest in deuterium and \hbox{$\rm{}^3{He}$},
since this observation implied a much larger
primordial (D+\hbox{$\rm{}^3{He}$})/H value than had been inferred from the
galactic observations.  More recently,
Rugers \& Hogan (1996b)\markcite{RugH96b}
measured $\rm D/H = 1.9^{+1.6}_{-0.9} \times 10^{-4}$ in another absorber of
Q0014+813 (at redshift $z=2.798$).  If the observed absorption lines do
indeed correspond to deuterium and not some coincidental interloper cloud
happening to lie at the corresponding velocity, then
this leads to correspondingly lower values of~$\eta \lesssim 1.7 \times
10^{-10}$ and~$\Omega_b \lesssim 0.006 h^{-2}$, close to the baryon density
$\Omega_b \approx 0.003$ observed in stars and gas, eliminating the need
for the existence of large amounts of baryonic dark matter.  These
observations also imply a decline by a factor of order~6 from the
primordial (D+\hbox{$\rm{}^3{He}$})/H value to the presolar value.  Galactic
chemical evolution models can obtain such a decline only if low mass stars
destroy \hbox{$\rm{}^3{He}$}, instead of creating it; even a {\it low\/}
primordial D/H value tends to result in excessive amounts
of~\hbox{$\rm{}^3{He}$} at the present epoch if stars
create~\hbox{$\rm{}^3{He}$} (see, e.g.,
Yang et al.\ 1984\markcite{Yang+84};
Walker et al.\ 1991\markcite{Wal+91};
Steigman \& Tosi 1992\markcite{SteT92};
Vangioni-Flam, Olive, \& Prantzos 1994\markcite{VangOP94};
Galli et al.\ 1994\markcite{Gal+94}, 1995\markcite{Gal+95})
--- note that a low value of
$\rm D/H = (2.3 \pm 0.3\> [stat] \pm 0.3\> [syst]) \times 10^{-5}$ has been
measured in an absorber at redshift $z=3.572$ of the quasar $1937-1009$
(Tytler \& Fan 1994\markcite{TytF94};
Tytler, Fan, \& Burles 1996\markcite{TytFB96}),
and that deuterium abundances in quasar absorbers are still being debated
(see, e.g.,
Tytler, Burles, \& Kirkman 1996\markcite{TytBK96};
Rugers \& Hogan 1997\markcite{RugH97};
Webb et al.\ 1997\markcite{Webb+97}).
Galli et al.\ (1994\markcite{Gal+94}, 1995\markcite{Gal+95})
suggested that a low-energy resonance in the
$\hbox{$\rm{}^3{He}$}+\hbox{$\rm{}^3{He}$}$
reaction could increase this rate sufficiently for low mass stars to
become net destroyers of \hbox{$\rm{}^3{He}$}.
Hogan (1995)\markcite{Hog95}
suggested a less speculative mechanism, namely, the {\it extra deep mixing\/}
below the conventional convective envelope on the
RGB, which is generally invoked to explain the anomalously
low \hbox{$\rm{}^{12}C$}/\hbox{$\rm{}^{13}C$} ratios observed in
low mass stars (below the \hbox{$\rm{}^{12}C$}/\hbox{$\rm{}^{13}C$}
values resulting from first dredge-up).

Over the last two decades,
the need for some such deep mixing process on the RGB has been
pointed out by many investigators, in order to understand the puzzle of the
low \hbox{$\rm{}^{12}C$}/\hbox{$\rm{}^{13}C$} ratios observed in low
mass Population~I RGB stars, and the \hbox{$\rm{}^{12}C$} depletion and
the [O/Fe] -- [Na/Fe] anticorrelation
discovered in low mass Population~II RGB stars (see, e.g.,
Dearborn, Eggleton, \& Schramm 1976\markcite{DeaES76};
Genova \& Schatzman 1979\markcite{GenS79};
Sweigart \& Mengel 1979\markcite{SweM79};
Gilroy 1989\markcite{Gil89};
Gilroy \& Brown 1991\markcite{GilB91};
Dearborn 1992\markcite{Dea92};
Smith \& Tout 1992\markcite{SmiT92};
Charbonnel 1994\markcite{Char94}, 1995a\markcite{Char95a},b\markcite{Char95b};
Boothroyd, Sackmann, \& Wasserburg 1995\markcite{BSW95}, hereafter BSW95;
Wasserburg, Boothroyd, \& Sackmann 1995\markcite{WBS95}, hereafter WBS95;
Denissenkov \& Weiss 1996\markcite{DenW96};
Charbonnel, Brown, \& Wallerstein 1998\markcite{CharBW98}).
Recently,
BSW95\markcite{BSW95}
and
WBS95\markcite{WBS95}
suggested that such deep
circulation might also occur in AGB stars of low mass, in order to account
for the anomalously low \hbox{$\rm{}^{18}O$} abundances observed in
these stars.

In this paper, we present consistent computations of
the effects of standard first and second dredge-up (in RGB and AGB stars,
respectively) on \hbox{$\rm{}^7{Li}$}
and~\hbox{$\rm{}^3{He}$}.  In addition, we have included
\hbox{$\rm{}^9{Be}$}, \hbox{$\rm{}^{10}B$},
and~\hbox{$\rm{}^{11}B$} in our calculations, since they are also
destroyed at relatively low temperatures in stars, and therefore can
provide additional signatures of stellar mixing.  For completeness, we
include the results for~\hbox{$\rm{}^4{He}$}.  We also present
calculations of the effects on these isotopes of ``{\it cool bottom
processing\/}''~(CBP)\hbox{}.  In this process, deep extra mixing
transports envelope material into the outer wing of the hydrogen-burning
shell, where it undergoes partial nuclear processing, and then transports
the material back out to the envelope.
We have modelled CBP by a deep circulation process, determining
some of the free parameters of our model
by requiring that our computations yield results that agree
with the observed RGB \hbox{$\rm{}^{12}C$}/\hbox{$\rm{}^{13}C$} ratios.
The CNO isotopes for these models are discussed in detail in
Boothroyd \& Sackmann (1998)\markcite{BS98}.


\section{Methods} \label{methods}

We considered stars of 29 different masses from 0.85 to~$9.0\>M_\odot$,
evolving them
self-consistently from the pre-main sequence through first dredge-up
up to either the helium core flash (which terminates the RGB for low mass
stars), or through second dredge-up to the first helium shell flash (for
intermediate mass stars and some low mass stars), or to core carbon
ignition during second dredge-up (for higher masses).
For evolutionary program details, see
Boothroyd \& Sackmann (1988)\markcite{BS88},
Sackmann, Boothroyd, \& Fowler (1990)\markcite{SBF90}, and
Sackmann, Boothroyd, \& Kraemer (1993)\markcite{SBK93}.

For solar metallicity, we used
a helium mass fraction $Y = 0.28$, with solar CNO abundances.
For lower
metallicities, we reduced~$Y$ (taking $\Delta Y / \Delta Z \approx 2$), and
increased the oxygen content, approximating the observed trend by
$\rm [O/Fe] \propto -0.5 [Fe/H]$ for $\rm [Fe/H] \ge -1$, and
constant $\rm [O/Fe] = +0.5$ for $\rm [Fe/H] < -1$ (see
Timmes, Woosley, \& Weaver 1995\markcite{TimWW95},
and references therein).
A fuller discussion of the CNO elements is given in
Boothroyd \& Sackmann (1998)\markcite{BS98}.
In addition to solar metallicity ($Z = 0.02$), we considered metallicities
appropriate to the Magellanic Clouds ($Z = 0.012$ and~$0.007$, with
$\rm [Fe/H] \approx -0.35$ and~$-0.7$, respectively),
an intermediate value of $Z = 0.003$
($\rm [Fe/H] \approx -1.2$), a Population~II metallicity of $Z = 0.001$
($\rm [Fe/H] \approx -1.7$), and an extreme Population~II metallicity of
$Z = 0.0001$ ($\rm [Fe/H] \approx -2.7$).  Since stars convert deuterium to
\hbox{$\rm{}^3{He}$} on the pre-main sequence, our program lumps deuterium
in with \hbox{$\rm{}^3{He}$};
our initial \hbox{$\rm{}^3{He}$} abundance is actually the sum of deuterium
and~\hbox{$\rm{}^3{He}$}.  Since the variation of (D+\hbox{$\rm{}^3{He}$})
as a function of time (or metallicity) is not well known, we generally chose
an initial ratio $\hbox{$\rm{}^3{He}/{}^4{He}$} = 4 \times 10^{-4}$ by
number (solar abundances), but also considered extreme values of this ratio,
namely, $4 \times 10^{-3}$ and $4 \times 10^{-5}$.
For solar metallicity we used a \hbox{$\rm{}^7{Li}$} abundance
$\hbox{$\log\,\varepsilon({}^7{\rm Li})$} = 3.0$, close to that of the present
interstellar medium
($\hbox{$\log\,\varepsilon({}^7{\rm Li})$}_{\rm ISM} \approx 3.3$); for
Population~II metallicities, we reduced the initial \hbox{$\rm{}^7{Li}$}
abundance to $\hbox{$\log\,\varepsilon({}^7{\rm Li})$} = 2.6$,
a factor of~5 below the present cosmic abundance.  Note that stellar models
which include microscopic diffusion of \hbox{$\rm{}^7{Li}$} suggest that
this is a reasonable initial value to match the observations of Spite plateau
lithium abundances in Population~II stars
(Vauclair \& Charbonnel 1995\markcite{VauC95}),
as do models which include rotational effects
(Pinsonneault, Deliyannis, \& Demarque 1992\markcite{PinDD92};
Deliyannis, Boesgaard, \& King 1995\markcite{DelBK95};
Vauclair \& Charbonnel 1995\markcite{VauC95});
it is hard to avoid some main sequence \hbox{$\rm{}^7{Li}$} depletion in these
stars.  Beryllium and boron isotopes were considered only for the
cases $Z = 0.02$ and $Z = 0.001$; we used solar abundances
($\rm Be/H = 1.3 \times 10^{-11}$, $\rm B/H = 4 \times 10^{-10}$,
and \hbox{$\rm{}^{11}B/{}^{10}B$}${} = 4.0$, by number:
Grevesse [1984]\markcite{Gre84}),
scaled linearly
with metallicity.  Note that initial abundances of the light elements are
generally not critical, as the depletion factors that we report in this
paper are insensitive to the absolute starting value.

Mass loss on the RGB and AGB was included via a
Reimers' (1975)\markcite{Rei75}
wind, with mass loss parameter $\eta \approx 0.6$, as discussed in
Boothroyd \& Sackmann (1998)\markcite{BS98}
(see also
Sackmann et al.\ 1993\markcite{SBK93});
values up to $\eta = 1.4$ for Population~I cases with masses $> 2.5\>M_\odot$
were tested, but the amount of mass lost was still insignificant at the
points where first and second dredge-up take place.
We used the OPAL 1995 interior opacities
(Iglesias \& Rogers 1996\markcite{IglR96}),
and Alexander molecular opacities
(Alexander \& Ferguson 1994\markcite{AlexF94})
at low temperatures; these latter require a value of $\alpha = 1.67$
(where $\alpha$ is the ratio of the convective mixing length to the
pressure scale height) in order to obtain a correct model of the Sun
(Sackmann et al.\ 1990\markcite{SBF90}, 1993\markcite{SBK93}).
Tests were made using older opacity tables.
Use of the interior opacities from the Los Alamos
Opacity Library (LAOL: from
Keady [1985]\markcite{Kea85})
yielded only slightly different amounts of dredge-up
(Boothroyd \& Sackmann 1998\markcite{BS98}),
while use of molecular opacities from
Sharp (1992)\markcite{Shar92}
required a value of $\alpha = 2.1$ but had no effect on dredge-up.
(Note that the value of~$\alpha$ has almost no effect on the depth of
dredge-up, as has already been noted by
Charbonnel [1994]\markcite{Char94}).
Nuclear reaction rates from
Caughlan~\& Fowler (1988)\markcite{CF88}
were used, except for the \hbox{$\rm{}^{17}O$}-destruction reactions
$\hbox{$\rm{}^{17}O$}(p,\alpha)\hbox{$\rm{}^{14}N$}$
and $\hbox{$\rm{}^{17}O$}(p,\gamma)\hbox{$\rm{}^{18}F$}
(\hbox{$e^{\hbox{$\scriptscriptstyle +$}}$}\nu)\hbox{$\rm{}^{18}O$}$,
where the rates of
Blackmon (1996)\markcite{Bl96}
(from measurements of
Blackmon et al.\ [1995]\markcite{Bla+95})
or of
Landr\'e et al.\ (1990)\markcite{La90}
were used.

For cool bottom processing (CBP), parametric computations were performed, with
envelope structures obtained from full evolution models of a $1\>M_\odot$
star in the appropriate stages of evolution.  Two types of models were
considered.  In the first type, as described in
WBS95\markcite{WBS95},
the star's structure was taken to be unchanging in time (``single
episode''); it was taken
at the point on the RGB where CBP is expected to begin (when the
hydrogen shell erases the composition discontinuity left behind by
first dredge-up).
The CBP was then computed over time period less than or comparable to the RGB
lifetime.
In the second type of model,
the change in envelope structure
as the star climbs the RGB was taken into account by interpolating between
full evolution models at 15 points on the~RGB (``evolving RGB'').
The CBP was assumed to start when the hydrogen shell erases the
composition discontinuity from first dredge-up, and continue until the tip
of the RGB was reached.
For these ``evolving RGB'' cases, metallicities $Z = 0.02$, 0.007,
0.001, and~0.0001 were considered (taking structures from the relevant
full evolution models), and higher masses were simulated by adding envelope
mass, and starting at the appropriate (higher-luminosity) RGB position
(this should be a good approximation, since the structure near the burning
shell on the RGB is almost independent of the envelope mass).
Isotopic abundances were taken from full evolutionary models, except for
\hbox{$\rm{}^7{Li}$}, where an additional main sequence depletion by a
factor of~50 was assumed for the $1\>M_\odot$ case (note that the observed
main sequence \hbox{$\rm{}^7{Li}$} depletion in low mass stars cannot be
predicted by standard stellar models).

In our two-stream ``conveyor-belt'' circulation model for CBP, matter from the
bottom of envelope convection streamed downward,
reaching a maximum temperature~$T_P$,
then returned upward and was mixed with the convective envelope (i.e., a
composition advection equation with nuclear burning, and no mixing
between downward and upward streams).  Envelope and
stream compositions were followed through time.  We assumed that the
temperature difference $\Delta\log\,T = \log\,T_P - \log\,T_H$ between
the bottom of mixing and the bottom of the hydrogen-burning shell was
constant, and treated it as a free parameter;
values selected for discussion were those satisfying the observational data,
as discussed in
WBS95\markcite{WBS95}
and
Boothroyd \& Sackmann (1998)\markcite{BS98}.
For a thin H-burning shell, assuming constant $\Delta\log\,T$ is roughly
equivalent to assuming extra mixing always reaches down to the same value
of the molecular weight gradient~$\nabla \mu$.

For CBP in the ``single episode'' circulation
models, $\Delta\log\,T \approx 0.17$ yielded a reasonable match to
the observed \hbox{$\rm{}^{12}C/{}^{13}C$} ratio on the RGB ($\sim 13 \pm 3$
for a $1.2 \> M_\odot$ star, or $\sim 11$ for a $1.0 \> M_\odot$ star: see
Gilroy [1989]\markcite{Gil89}),
after a processing
time $t_{\rm mix:RGB} \approx 1.25 \times 10^7\>$yr (see Fig.~\ref{figrgbt}).
During this time, the star's luminosity increases by only~60\%
(Sackmann et al.\ 1993\markcite{SBK93}),
so the assumption of static envelope structure is not unreasonable (the
luminosity increase is by a factor of~100 at the tip of the RGB, after a time
$\tau_{\scriptscriptstyle\!\rm RGB} \sim 7 \times 10^7\>$yr).
%
For CBP in the ``evolving RGB'' circulation models,
$\Delta\log\,T = 0.262 \pm 0.010$ matched the above observed carbon isotope
ratios on the RGB (Fig.~\ref{figrgbevol}a; see also
Boothroyd \& Sackmann 1998\markcite{BS98}).

The other key free parameter was the stream mass flow rate~$\dot M_p$.
This must be slower than that of convection in RGB or AGB envelopes
($\dot M_p \ll \dot M_{\rm conv} \sim 1\>M_\odot$/yr),
while the streams must move faster than the speed with which the
H-shell burns its way outward ($\dot M_p \gtrsim \dot M_c$).
We explored a wide range of $\dot M_p$~values.
For equal downward and upward velocities, the fractional
areas at the base of the convective envelope occupied by downward and
upward streams are equal.  The downward stream then spends a time
$\Delta t_d = 0.5 \, \Delta M_r / \dot M_p$ in a layer~$\Delta M_r$,
the upward stream spending the same time $\Delta t_u$ there on its way
out.  When downward and upward velocities (and respective fractional
areas $f_d$ and~$f_u$) are {\it not\/} equal, these times become
$\Delta t_d = f_d \Delta M_r / \dot M_p$ and $\Delta
t_u = f_u \Delta M_r / \dot M_p$.
The total time spent in any mass layer
is independent of $f_d$ and~$f_u$, for $f_d + f_u = 1$; thus the total
amount of nuclear burning is generally independent of~$f_u/f_d$.
The \hbox{$\rm{}^7{Li}$} production via the Cameron-Fowler mechanism
(Cameron 1955\markcite{Cam55};
Cameron \& Fowler 1971\markcite{CamF71})
is an exception; it {\it does\/} depend on~$f_u$, which
affects the speed with which \hbox{$\rm{}^7{Be}$} is transported
upwards, and thus the amount which survives to reach cool layers before
decaying into~\hbox{$\rm{}^7{Li}$}.  Let $f \equiv f_u/f_d$; besides
$f = 1$ (i.e., $f_d = f_u = 0.5$), we considered $f = 9$ and~99.


\section{Results} \label{results}

We present first and second dredge-up results from standard stellar models,
and cool bottom processing (CBP) results from our ``single episode'' and
``evolving~RGB'' circulation models of extra mixing.  Note that some results
for \hbox{$\rm{}^4He$}, \hbox{$\rm{}^3He$}, and the CNO isotopes are
tabulated in
Boothroyd \& Sackmann (1998)\markcite{BS98};
more detailed tables are available from the authors\footnote[1]{\ Send
e-mail to
{\tt aib}@{\tt krl.caltech.edu} to obtain tables of results; also available
at A.~I.~B.'s Web page: {\tt http://www.krl.caltech.edu/${}^{\sim}$aib/}}.


\subsection{Standard First and Second Dredge-up} \label{standr}


\begin{figure}[!t]
     \bootplotfidtwo{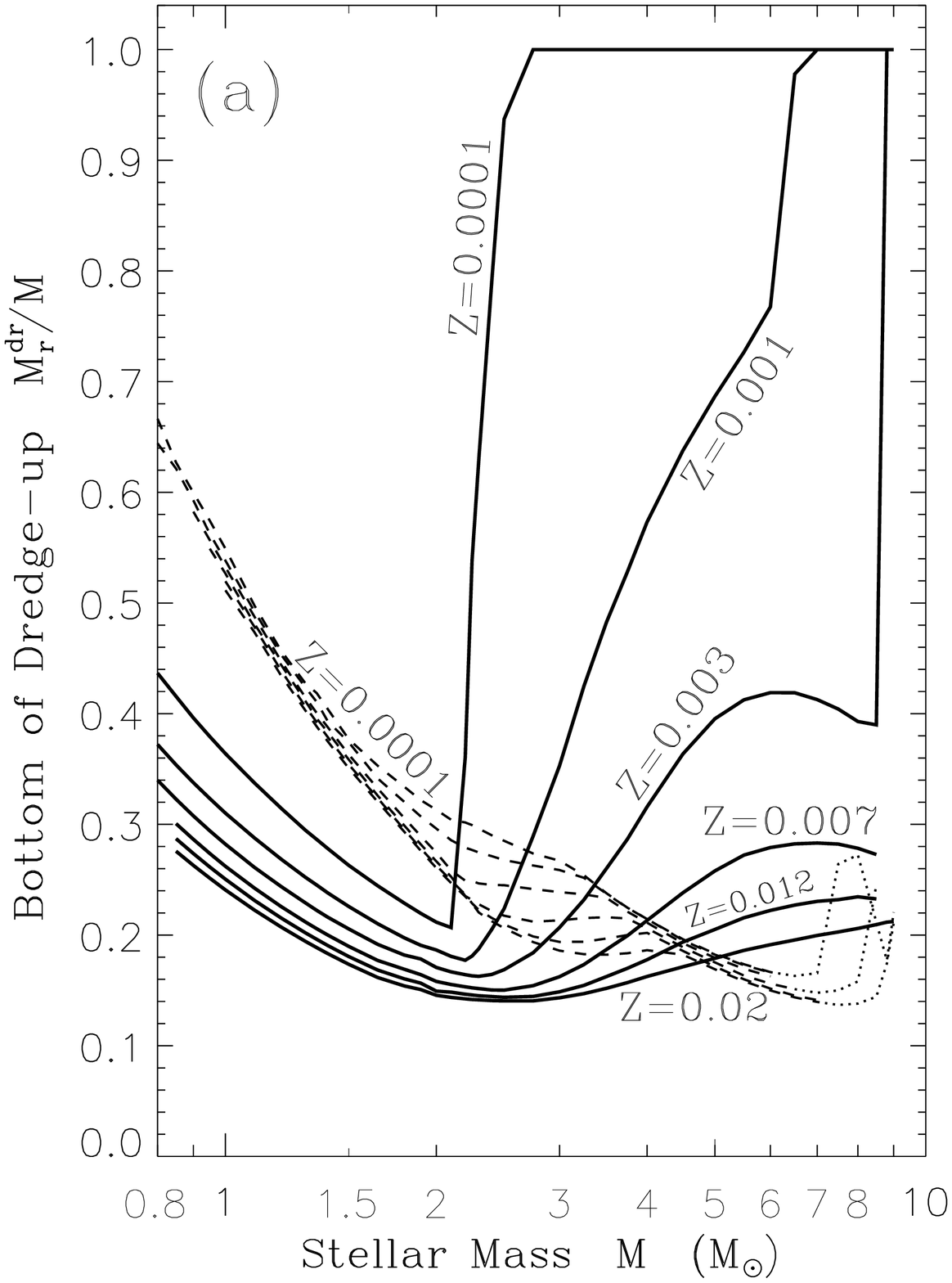}{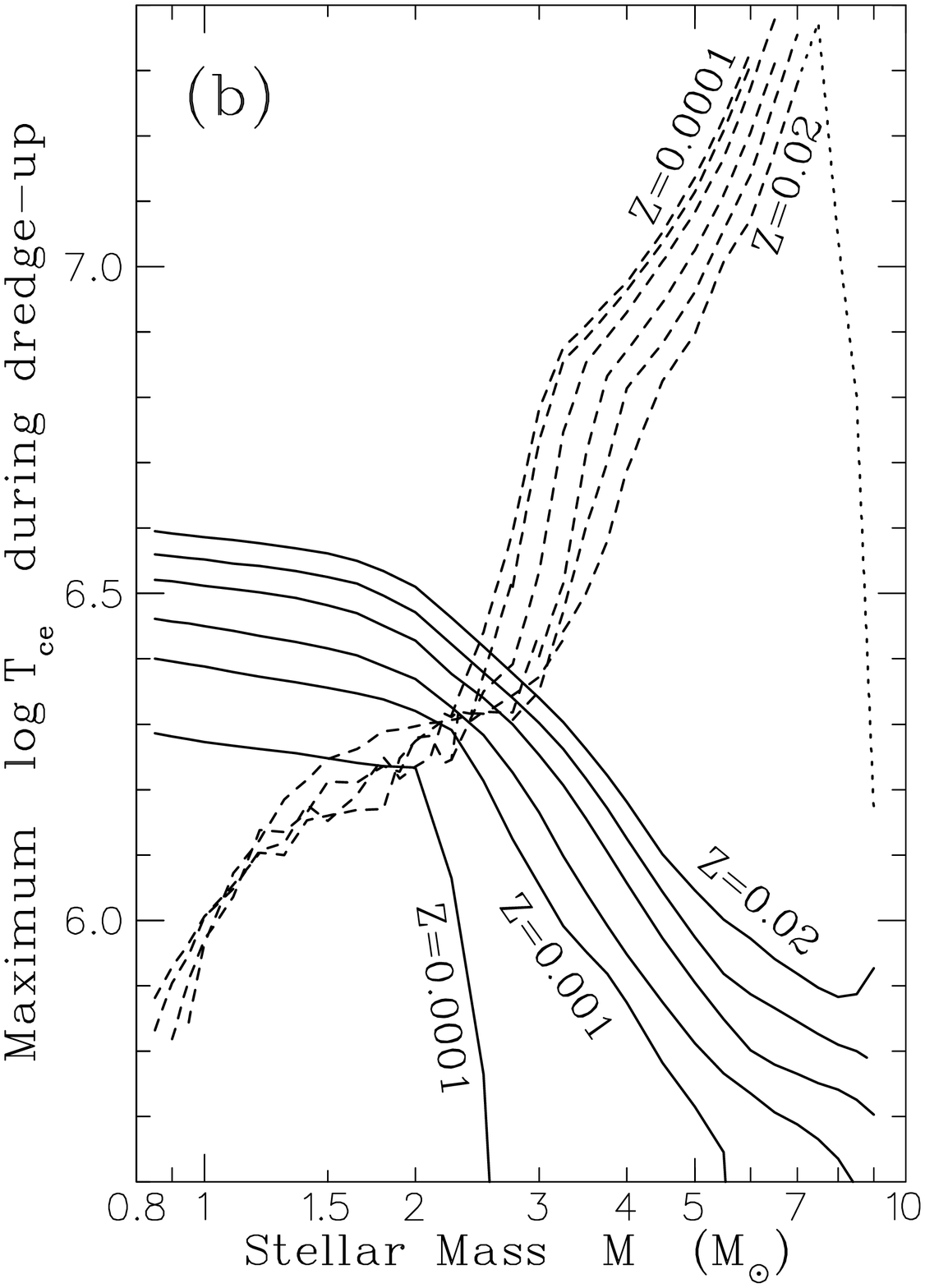}{4 true
       in}{48}{-183}{-40}{46}{-100}{-34}

\caption{(a)~Maximum depth in mass of the convective
envelope~($M_r^{\rm dr}$),
relative to the initial stellar mass~$M$, of first and second dredge-up
({\it solid\/} and {\it dashed curves}, respectively) as a function of
stellar mass~$M$.
The metallicities are indicated on the curves for first dredge-up; the
second dredge-up curves are in the same order.
For $Z = 0.02$, 0.007, and~0.0001, the {\it dotted\/}
continuations of the dashed curves show the depth reached by second
dredge-up at the point where the program
failed during core carbon ignition.
(b)~The maximum temperature~$T_{\rm ce}$ at the bottom of the
convective envelope during first and second dredge-up ({\it solid\/} and
{\it dashed curves}, respectively) for the same metallicities as~(a);
for $Z = 0.02$, the value of~$T_{\rm ce}$
is also shown at the time of core carbon ignition, for stars of
mass $> 7\>M_\odot$ ({\it dotted\/} continuation of dashed curve).}

 \label{figdrdep}

\end{figure}
\placefigure{figdrdep}

Figure~\ref{figdrdep}a presents the mass layer~$M_r/M$ down to
which the convective envelope reaches, at its deepest penetration
during standard first and second dredge-up (for stars of $> 7\>M_\odot$,
the depth of second dredge-up is shown at the moment when the program
failed during core carbon ignition).
Note that we define ``second dredge-up'' in low mass stars as the stage
of deepest convective penetration on the early AGB, even if this is
shallower than first dredge-up.
Figure~\ref{figdrdep}b displays the corresponding
temperatures~$T_{\rm ce}$ at the base of the convective envelope.
For $Z = 0.02$ stars of mass $\le 1\>M_\odot$, there is some burning
of~\hbox{$\rm{}^7{Li}$} during first dredge-up, when $T_{\rm ce}$
reaches nearly $4 \times 10^6\>$K\hbox{}.
Higher temperatures are attained during second dredge-up in intermediate
mass stars, but no
\hbox{$\rm{}^7{Li}$} burning takes place in any but the highest
mass stars ($\sim 6 - 7\>M_\odot$), as the timescales are too short.


\begin{figure}[!t]
  \plottwo{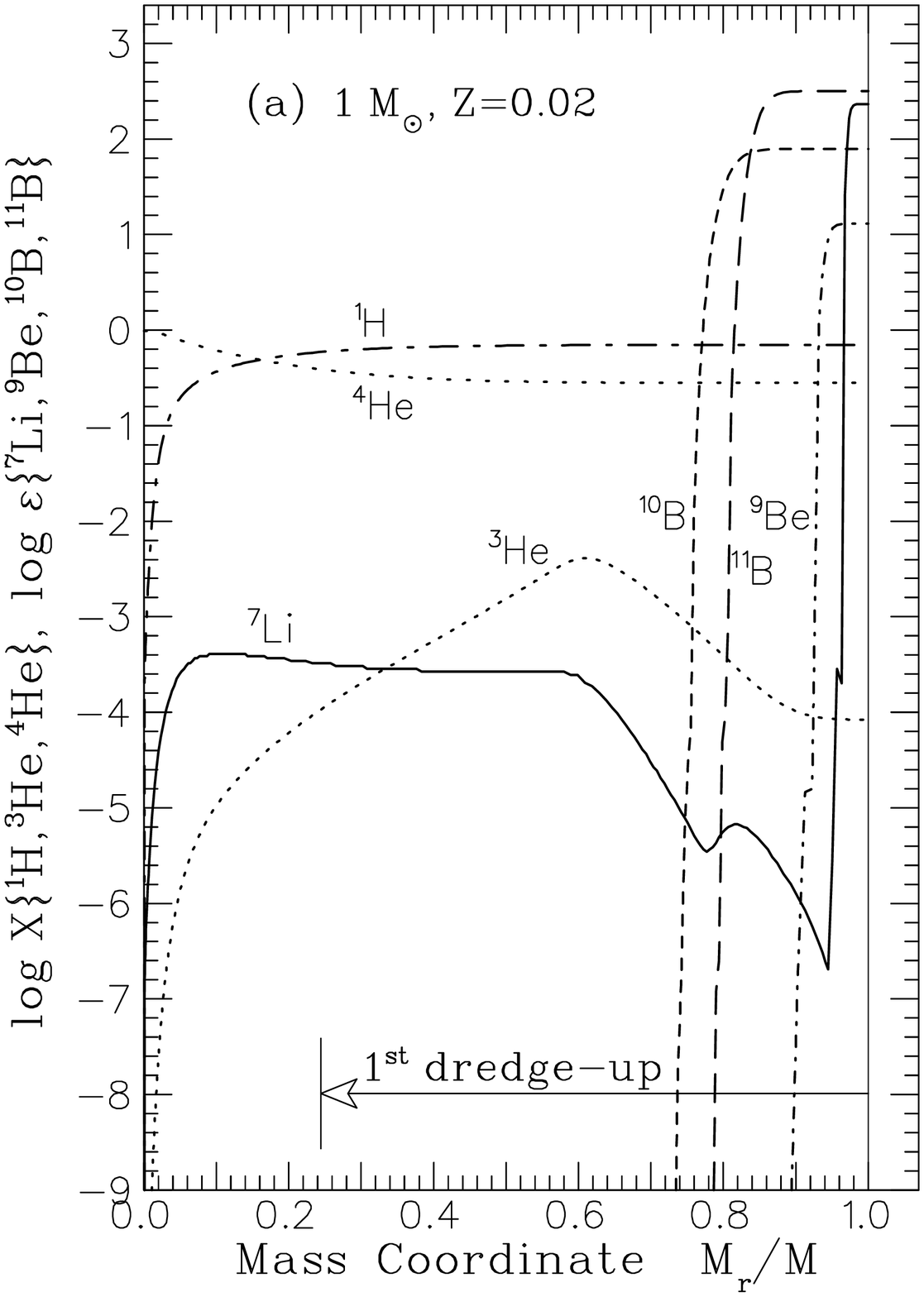}{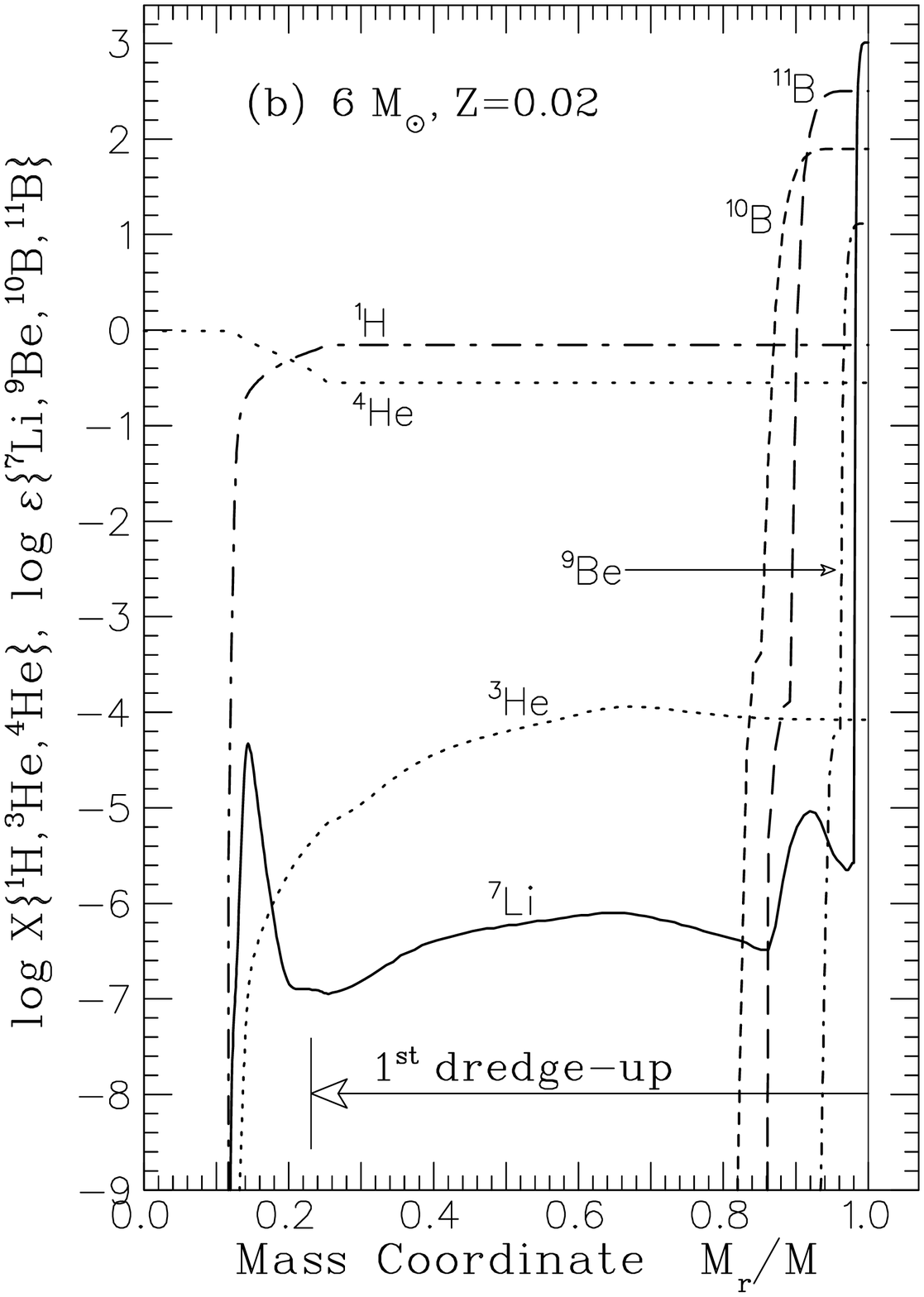}

\caption{The abundance profiles of the light elements as a function of
the mass coordinate~$M_r/M$ at the end of the main sequence for $Z = 0.02$
stars (a)~of $1\>M_\odot$, and (b)~of $6\>M_\odot$.
The depth of standard first dredge-up on the RGB is shown by the arrows.}

 \label{figprof}

\end{figure}
\placefigure{figprof}

Figure~\ref{figprof} illustrates where in a star's envelope
\hbox{$\rm{}^7{Li}$}, \hbox{$\rm{}^9{Be}$}, \hbox{$\rm{}^{10}B$},
and~\hbox{$\rm{}^{11}B$} are destroyed, and where in the interior
the pocket of \hbox{$\rm{}^3{He}$}
has been created, prior to first dredge-up.  One sees that
\hbox{$\rm{}^7{Li}$} is burned up first, followed by
\hbox{$\rm{}^9{Be}$}; further in, \hbox{$\rm{}^{11}B$}
and~\hbox{$\rm{}^{10}B$} burn at similar depths
(\hbox{$\rm{}^{11}B$}~slightly
before~\hbox{$\rm{}^{10}B$}).  The pocket of~\hbox{$\rm{}^3{He}$} peaks
considerably deeper in the interior.
First dredge-up will dilute the pockets of \hbox{$\rm{}^7{Li}$},
\hbox{$\rm{}^9{Be}$}, \hbox{$\rm{}^{11}B$}, and~\hbox{$\rm{}^{10}B$},
and will enrich the surface with \hbox{$\rm{}^3{He}$} from the pocket
below.  Table~\ref{tblpocsiz} illustrates the depths of these pockets in
stars of different masses and metallicities, prior both to first dredge-up
on the RGB and to second dredge-up on the~AGB\hbox{}.


\begin{table*}[!b]

\caption{Size of the Pockets of \hbox{$\rm{}^7{Li}$}, \hbox{$\rm{}^9{Be}$},
 \hbox{$\rm{}^{10}B$}, and \hbox{$\rm{}^{11}B$} Prior to First and
 Second Dredge-up$\,$\tablenotemark{a}}

  \label{tblpocsiz}

\begin{center}
\tabcolsep=0.3 em
\begin{tabular*}{\hsize}{@{}@{\extracolsep{\fill}}{l}*{9}{c}@{}}
\tableline
\tableline
     \noalign{\smallskip}
 $Z\,$ \& & $M$ & \multicolumn{7}{c}{\dotfill $(M-M_r)/M$\dotfill} & \\
 stage & $(M_\odot)$ & dr$\,$\tablenotemark{b} & \hbox{$\rm{}^7{Li}$} &
  \hbox{$\rm{}^9{Be}$} & \hbox{$\rm{}^{10}B$} & \hbox{$\rm{}^{11}B$} &
  \hbox{$\rm{}^3{He}$} & ${\hbox{$\rm{}^3{He}$}_p}\,$\tablenotemark{c} &
  $\smash{\raise7pt\hbox{$\displaystyle {\hbox{$\rm{}^3{He}$}_p
   \, \tablenotemark{d} \! \over \hbox{$\rm{}^3{He}$}_s}$}}$  \\
     \noalign{\smallskip}
\tableline
     \noalign{\smallskip}
 0.02, & 1.0 & 0.756 & 0.037 & 0.060 & 0.194 & 0.152 & 0.826 & 0.391 & 49. \\
 \ RGB & 2.5 & 0.858 & 0.0148 & 0.0312 & 0.1172 & 0.0872 & 0.777 & 0.369 &
   5.8 \\
 & 6.0 & 0.769 &  0.0132 & 0.0275 & 0.1055 & 0.0770 & 0.604 & 0.327 & 1.37 \\
\noalign{\bigskip}
 0.001, & 1.0 & 0.689 & 0.024 & 0.055 & 0.176 & 0.139 & 0.724 & 0.336 & 54. \\
 \ RGB & 2.5 & 0.746 & 0.0104 & 0.0240 & 0.0932 & 0.0696 & 0.751 & 0.283 &
   8.0 \\
 & 6.0 & 0.0008 & 0.0080 & 0.0205 & 0.0788 & 0.0575 & 0.510 & 0.260 & 1.69 \\
\noalign{\bigskip}
 0.02, & 2.5 & 0.788 & 0.669 & 0.717 & 0.760 & 0.753 & 0.783 & 0.778 & 1.10 \\
 \ AGB & 6.0 & 0.845 & 0.559 & 0.647 & 0.708 & 0.691 & 0.754 & 0.741 & 1.01 \\
\noalign{\bigskip}
 0.001, & 2.5 & 0.729 & 0.255 & 0.324 & 0.508 & 0.464 & 0.659 & 0.627 & 1.08 \\
 \ AGB & 6.0 & 0.837 & 0.0079 & 0.0204 & 0.0787 & 0.0574 & 0.512 & 0.260 &
   1.68 \\
     \noalign{\smallskip}
\tableline
     \noalign{\vskip -30.7pt}
\end{tabular*}
\end{center}

\tablenotetext{a}{\ Bottom of pocket is defined as point where abundance
drops by a factor of~2.}

\tablenotetext{b}{\ Fraction of the star's mass contained in the convective
envelope at the time of deepest dredge-up.}

\tablenotetext{c}{\ Depth of the \hbox{$\rm{}^3{He}$} peak.}

\tablenotetext{d}{\ Peak \hbox{$\rm{}^3{He}$} abundance, relative to its
surface abundance.  Note that the peak subsequently grows somewhat higher
than the values quoted here, before it is actually dredged up.}

\end{table*}
\placetable{tblpocsiz}


\begin{figure}[!t]
  \plotfiddle{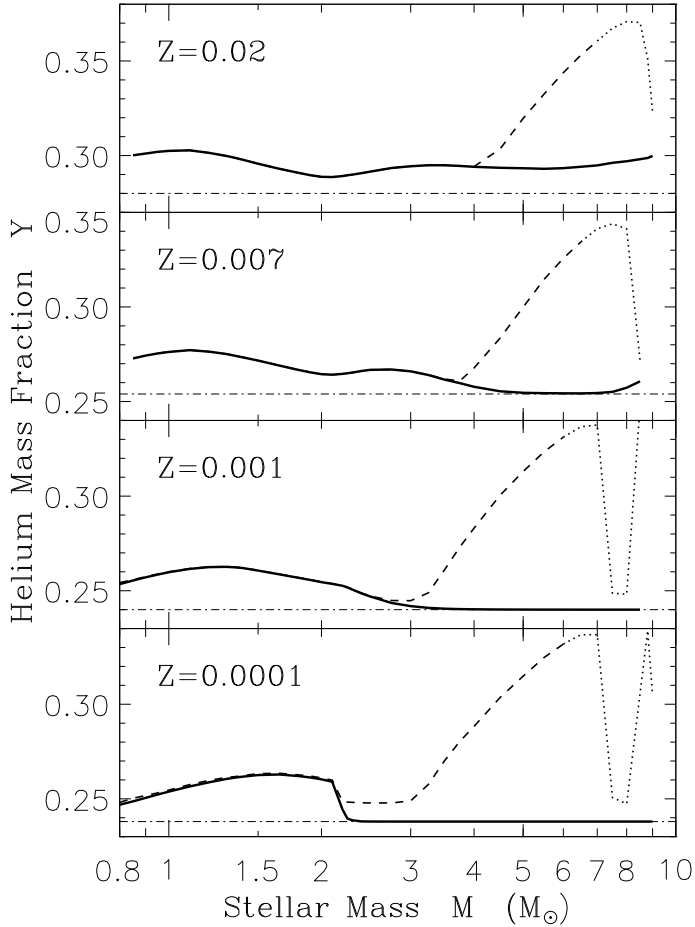}{4.5 true in}{0}{54}{54}{-180}{-43}

\caption{Envelope \hbox{$\rm{}^4{He}$} mass fraction~$Y$ due to standard
first and second dredge-up ({\it solid\/} and {\it dashed curves},
respectively), or at core carbon ignition during second dredge-up
({\it dotted\/} continuation of
dashed curves), as a function of initial stellar mass~$M$ for various input
metallicities~$Z$.  The initial stellar abundance is given by the
{\it dot-dashed lines}.}

 \label{fighe4}

\end{figure}
\placefigure{fighe4}

Figure~\ref{fighe4} displays the envelope \hbox{$\rm{}^4{He}$} enrichment
produced by first and second dredge-up.  There is little
metallicity dependence; first dredge-up in
low mass stars increases the helium mass fraction
by~$\Delta Y \sim 0.02$, while second dredge-up in $\sim 7\>M_\odot$ stars
yields $\Delta Y \sim 0.1$.  One can compute the contribution of these
stars to the \hbox{$\rm{}^4{He}$} enrichment of the interstellar medium
(similar estimates for CNO isotopes are discussed in more detail in
Boothroyd \& Sackmann [1998]\markcite{BS98}).
Using the initial--final mass relationship
(Weidemann \& Koester 1983\markcite{WeidK83};
Weidemann 1984\markcite{Weid84}),
one can estimate the envelope mass ejected
for a given initial stellar mass.  Multiplying mass ejected
by the mass fraction of \hbox{$\rm{}^4{He}$} from Figure~\ref{fighe4},
and weighting the results by an initial mass function
$\phi(M) \propto M^{-2.3}$
(Salpeter 1955\markcite{Sal55}),
one can estimate the contribution to the interstellar medium.  Comparing
with similar \hbox{$\rm{}^4{He}$} computations for supernovae
(Weaver \& Woosley 1993\markcite{WeaW93}),
one finds that low and intermediate mass stars ($1 - 12\>M_\odot$)
inject nearly as much \hbox{$\rm{}^4{He}$} into the interstellar
medium as do supernovae.


\begin{figure}[!t]
     \bootplotfidtwo{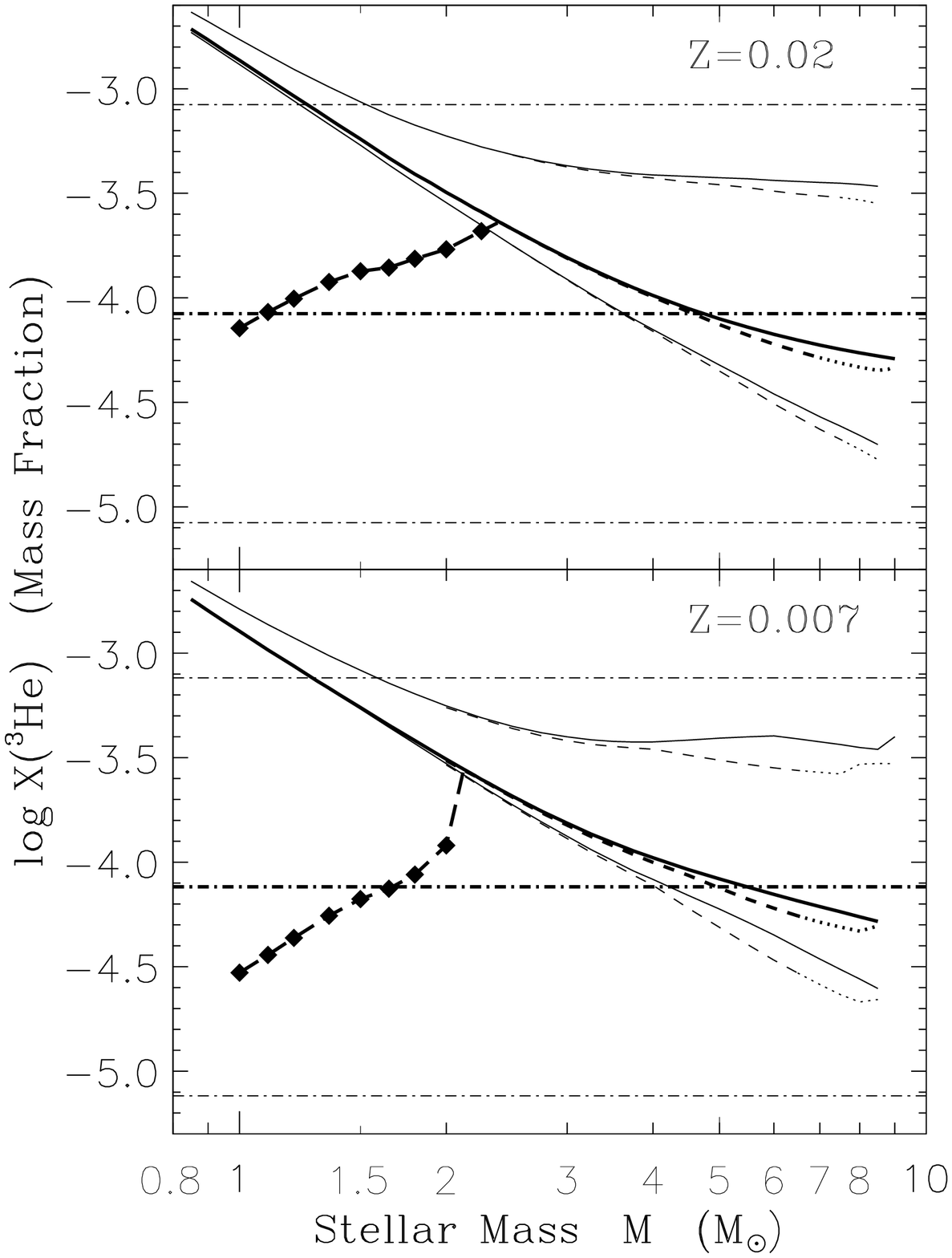}{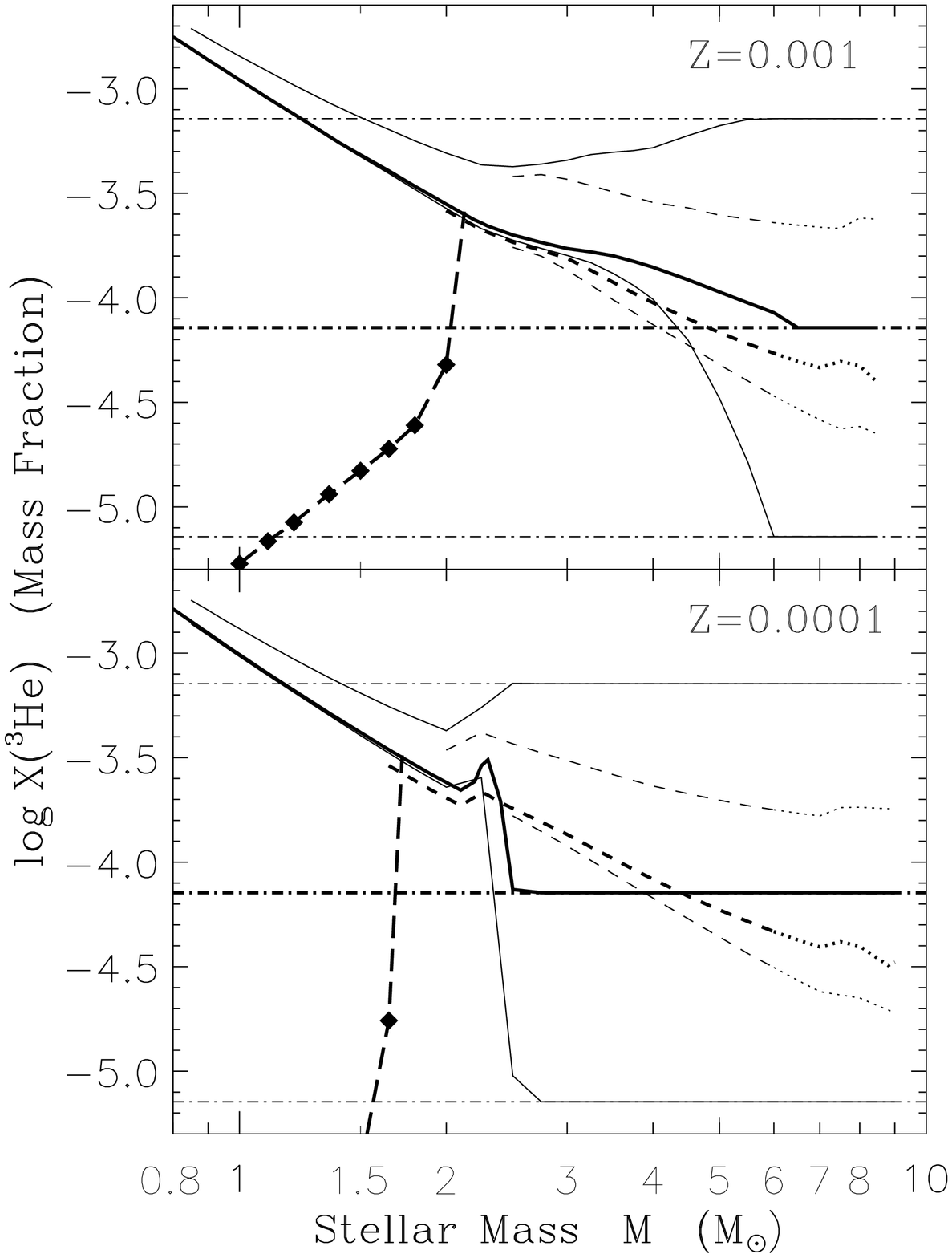}{4 true
       in}{48}{-180}{-40}{48}{-105}{-40}

\caption{Envelope~\hbox{$\rm{}^3{He}$} abundance due to standard
first and second dredge-up ({\it solid\/} and {\it dashed curves},
respectively), or at core carbon ignition ({\it dotted\/} continuation of
dashed curves); {\it dot-dashed lines\/} show initial
stellar~\hbox{$\rm{}^3{He}$} abundances.  Heavy (central)
curves are for our standard
initial ratio $\hbox{$\rm{}^3{He}/{}^4{He}$} = 4 \times 10^{-4}$ by
number; light curves are for extreme cases $4 \times 10^{-3}$ and
$4 \times 10^{-5}$.  {\it Diamonds\/} connected by
{\it long-dashed curves\/} show the effect of cool bottom processing (CBP)
on the RGB, from our ``evolving RGB'' CBP models.}

 \label{fighe3}

\end{figure}
\placefigure{fighe3}

Figure~\ref{fighe3} is a key diagram.  It illustrates the conventional
viewpoint that low mass stars ($M \lesssim 4\>M_\odot$)
should be a considerable source of \hbox{$\rm{}^3{He}$} in the universe,
according to the predictions of standard dredge-up theory.
Note that the first and second dredge-up results of Figure~\ref{fighe3}
are in good agreement ($\sim 10$\%) with the low mass star results of
Charbonnel (1995a\markcite{Char95a},b\markcite{Char95b}),
the low and intermediate mass star results of
Dearborn, Steigman, \& Tosi (1996)\markcite{DeaST96},
and the intermediate mass star results of
Weiss, Wagenhuber, \& Denissenkov (1996)\markcite{WeisWD96}
(though the latter find $\sim 30$\% less \hbox{$\rm{}^3{He}$} dredge-up
than the other authors, for stellar masses below~$1.5\>M_\odot$).
Figure~\ref{fighe3} also illustrates the \hbox{$\rm{}^3{He}$} depletion that
is expected due to extra mixing in low mass stars, as computed via our
``evolving RGB'' CBP models ({\it diamonds\/} on
{\it long-dashed curves}).  For low enough
stellar masses, overall depletion of \hbox{$\rm{}^3{He}$} is obtained,
relative to its initial abundance.  On average, one finds that stars are
{\it net destroyers\/} of~\hbox{$\rm{}^3{He}$} in the universe, particularly
low-metallicity stars; this is discussed below in~\S~\ref{cbp}.
(Note that Fig.~\ref{fighe3} does {\it not\/} show the effect of hot bottom
burning in intermediate mass AGB stars, which would destroy almost all the
\hbox{$\rm{}^3{He}$} in Population~I stars of $4 - 7\>M_\odot$, and
in Population~II stars of $3.5 - 6\>M_\odot$
[Boothroyd \& Sackmann 1992\markcite{BS92};
Sackmann \& Boothroyd 1998\markcite{SB98}].)


\begin{figure}[!t]
  \plottwo{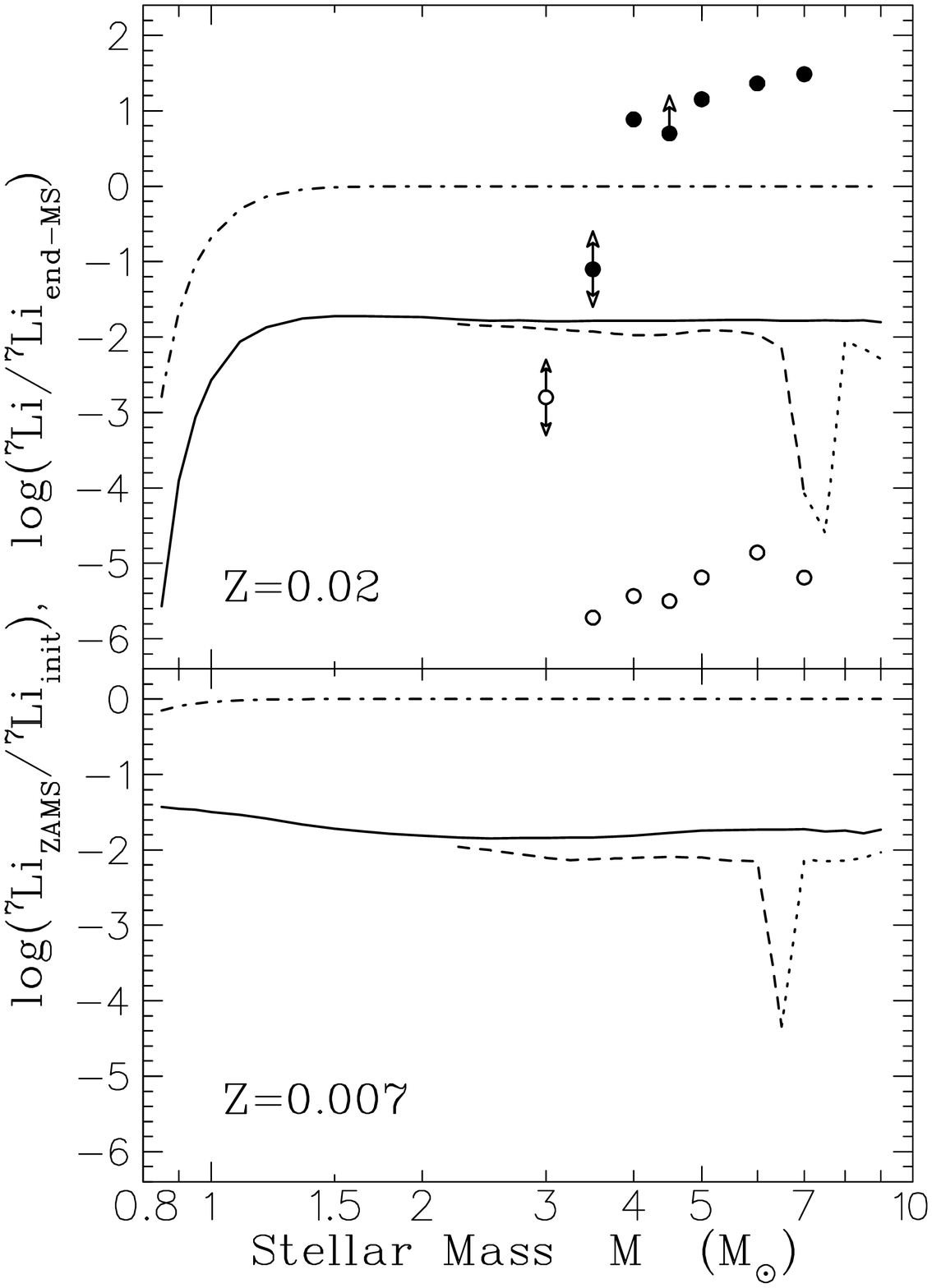}{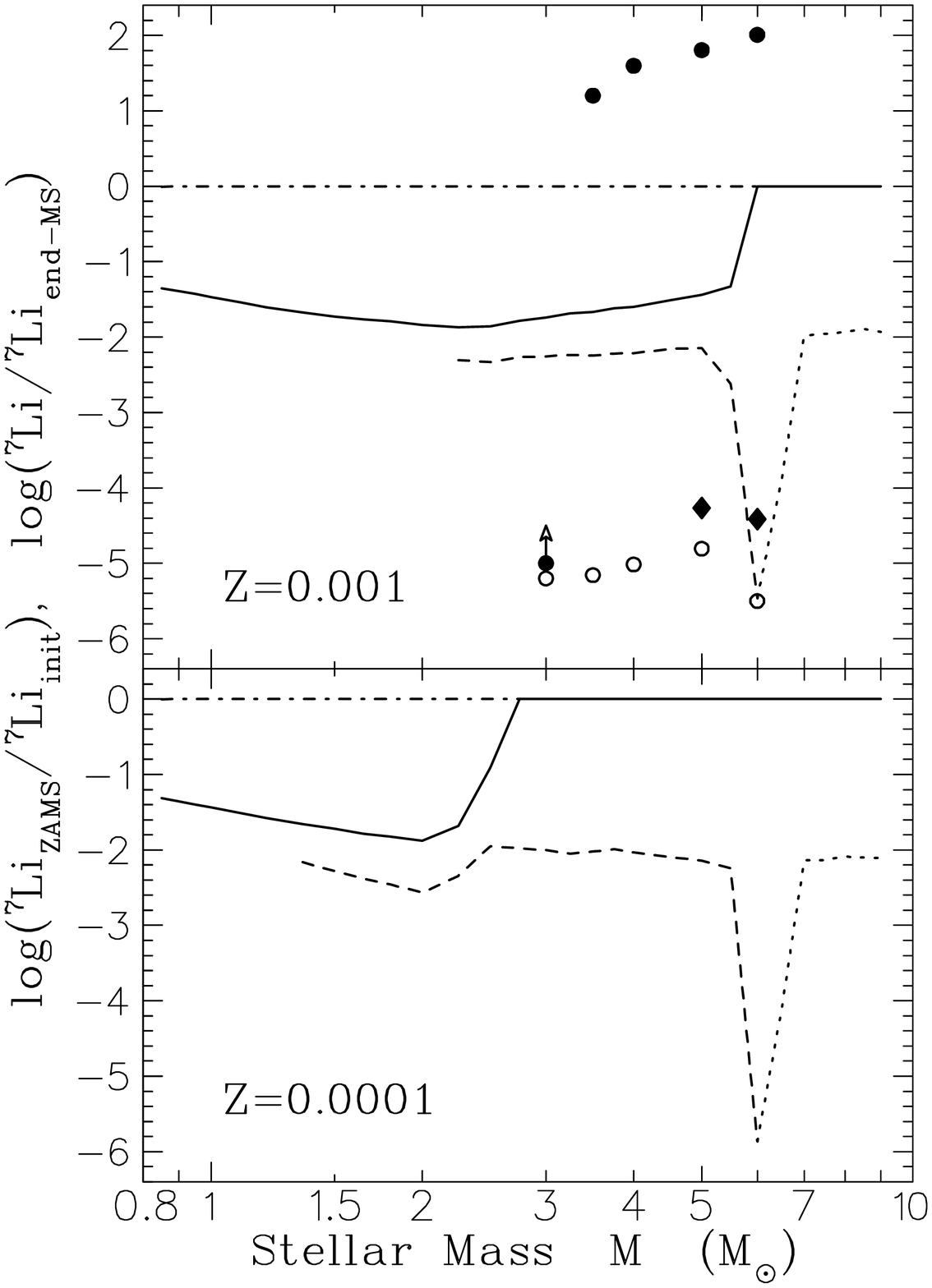}

\caption{Envelope \hbox{$\rm{}^7{Li}$} abundance changes due to standard
first and second dredge-up ({\it solid\/} and {\it dashed curves},
respectively), or at core carbon ignition ({\it dotted\/} continuation of
dashed curves), relative to the values at the end of the main sequence
(depletion during the main sequence is not shown).
{\it Dot-dashed curves\/} show the pre-main sequence \hbox{$\rm{}^7{Li}$}
burning, relative to the initial (interstellar medium) value.  Separated
symbols show the effect of AGB hot bottom burning for $Z = 0.02$ and~0.001:
the minimum AGB \hbox{$\rm{}^7{Li}$} abundance at the start of hot bottom
burning ({\it open circles\/}), peak AGB \hbox{$\rm{}^7{Li}$} abundance
({\it solid circles\/}),
and final equilibrium hot bottom burning \hbox{$\rm{}^7{Li}$} abundances
({\it solid diamonds\/} --- note that this equilibrium may not be
reached by Population~I stars and lower
mass Population~II stars).  Arrows indicate directions of possible shifts
in values, as expected from variations in AGB mass loss rates, or from
continuation of an incomplete model run (see
Sackmann \& Boothroyd 1995\protect\markcite{SB95},
1998\protect\markcite{SB98}).}

 \label{figli7}

\end{figure}
\placefigure{figli7}

Figure~\ref{figli7} summarizes the reduction of the surface
\hbox{$\rm{}^7{Li}$} abundance due to pre-main sequence
burning and first and second dredge-up.  Pre-main sequence burning
({\it dot-dashed curve} in Fig.~\ref{figli7}) is important only for masses
$\lesssim 1.2\>M_\odot$ at near-solar metallicity; there is no significant
pre-main sequence \hbox{$\rm{}^7{Li}$} burning in Population~II stars
of~$\ge 0.85\>M_\odot$.
Possible effects of main-sequence \hbox{$\rm{}^7{Li}$} depletion on the
first dredge-up depletion factors of Figure~\ref{figli7} are discussed below.

The \hbox{$\rm{}^7{Li}$} depletion due to first and second dredge-up
is usually due entirely to dilution (by a factor of~$\sim 60$) of the
surface lithium pocket, and this dilution is relatively insensitive
to the stellar mass and metallicity
(varying by less than a factor of~3).  In addition to dilution, however, there
are some cases where \hbox{$\rm{}^7{Li}$} also burns during dredge-up, at the
bottom of the (standard) deep convective envelope.
This occurs during first dredge-up only for Population~I stars of
masses $\le 1\>M_\odot$, and during second dredge-up only for Population~I
stars of $\sim 7\>M_\odot$ and Population~II stars of $\sim 6\>M_\odot$.

Extra mixing on the main sequence can destroy lithium there; one may question
whether this would affect the first dredge-up dilution factors.
%
%
Low mass stars are observed to experience considerable \hbox{$\rm{}^7{Li}$}
depletion on the main sequence (see, e.g.,
Hobbs \& Pilachowski 1988\markcite{HobP88},
and references therein).
This will affect the RGB abundances, but not the dilution factor (since the
{\it size\/} of the lithium pocket on the main sequence cannot be reduced
significantly --- it does not extend much below the main sequence convective
envelope, at least for stars $\lesssim 1 \> M_\odot$).
Stars of mass $\gtrsim 1.4\>M_\odot$
do not exhibit reduced {\it surface\/} lithium abundances while
on the main sequence (see, e.g.,
Boesgaard \& Tripicco 1987\markcite{BoeT87};
Balachandran 1991\markcite{Bal91}).
However, there may be some evidence that extra mixing can cause additional
lithium depletion in the {\it interior}, reducing the size of the lithium
pocket (and thus increasing the first dredge-up dilution factor).
For the Hyades, there is no such evidence:
the lithium abundances on the RGB are completely explainable
by first dredge-up dilution alone, as pointed out by
Duncan et al.\ (1998)\markcite{Dun+98}.
%
Possible evidence for additional lithium depletion comes from
Gilroy's (1989)\markcite{Gil89}
observations of \hbox{$\rm{}^7{Li}$} in open cluster giants of intermediate
mass; non-LTE corrections to the observed lithium abundances
(Carlsson et al.\ 1994\markcite{Carl+94})
may reduce the effect, but probably will not eliminate it.
%
%

Figure~\ref{figli7} also summarizes the \hbox{$\rm{}^7{Li}$} creation
in AGB stars of $\sim 4 - 7\>M_\odot$ due to hot bottom burning
(Sackmann \& Boothroyd 1992\markcite{SB92}, 1995\markcite{SB95},
1998\markcite{SB98}).
First, \hbox{$\rm{}^7{Li}$} is
burned, being depleted by several orders of magnitude.  As hot bottom
burning becomes stronger, the Cameron-Fowler mechanism
(Cameron 1955\markcite{Cam55};
Cameron \& Fowler 1971\markcite{CamF71})
yields a large increase in the envelope \hbox{$\rm{}^7{Li}$} abundance.
Independent of past \hbox{$\rm{}^7{Li}$} history,
such super-rich lithium stars attain peak \hbox{$\rm{}^7{Li}$}
abundances $10 - 20$ times the present interstellar \hbox{$\rm{}^7{Li}$}
abundance, and may contribute significant amounts of
\hbox{$\rm{}^7{Li}$} to the interstellar medium
(Sackmann \& Boothroyd 1995\markcite{SB95}, 1998\markcite{SB98}).
The envelope \hbox{$\rm{}^7{Li}$}
is produced from~\hbox{$\rm{}^3{He}$}; as \hbox{$\rm{}^3{He}$} is burned,
the \hbox{$\rm{}^7{Li}$} abundance declines again.  If hot
bottom burning continues long enough, \hbox{$\rm{}^3{He}$}~destruction
reaches equilibrium with its creation via the $p$-$p$ chain, and
the \hbox{$\rm{}^7{Li}$} abundance reaches a
final value $4 - 5$ orders of magnitude below the interstellar abundance
(see 5 and $6\>M_\odot$ $Z = 0.001$
cases in Fig.~\ref{figli7}); stellar runs for
Population~I stars were terminated before this stage was reached, since they
approach this final equilibrium much more slowly than Population~II stars
(and may in fact never reach it at all, depending on the point of onset of
the AGB ``superwind'').


\begin{figure}[!t]
  \plottwo{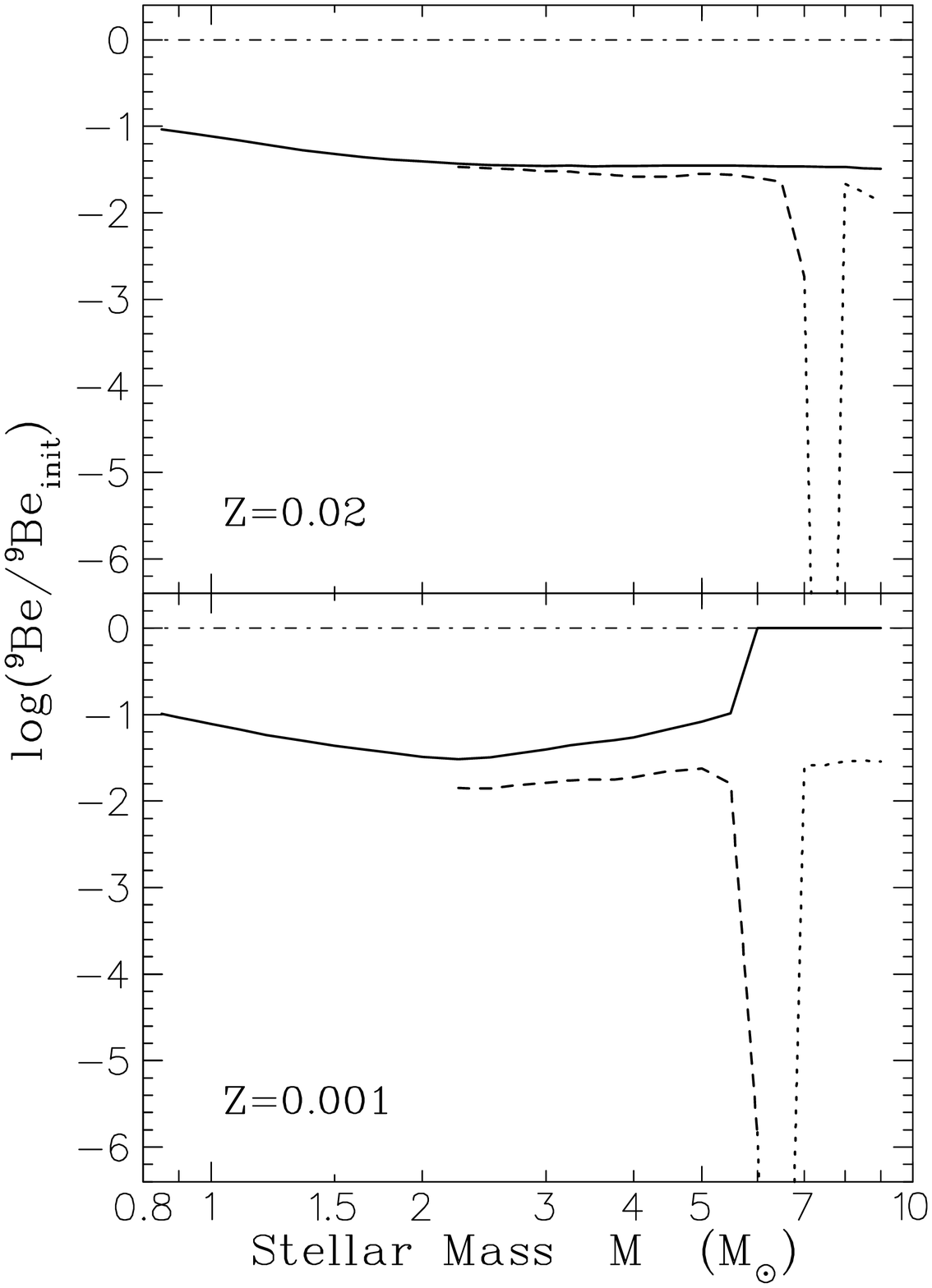}{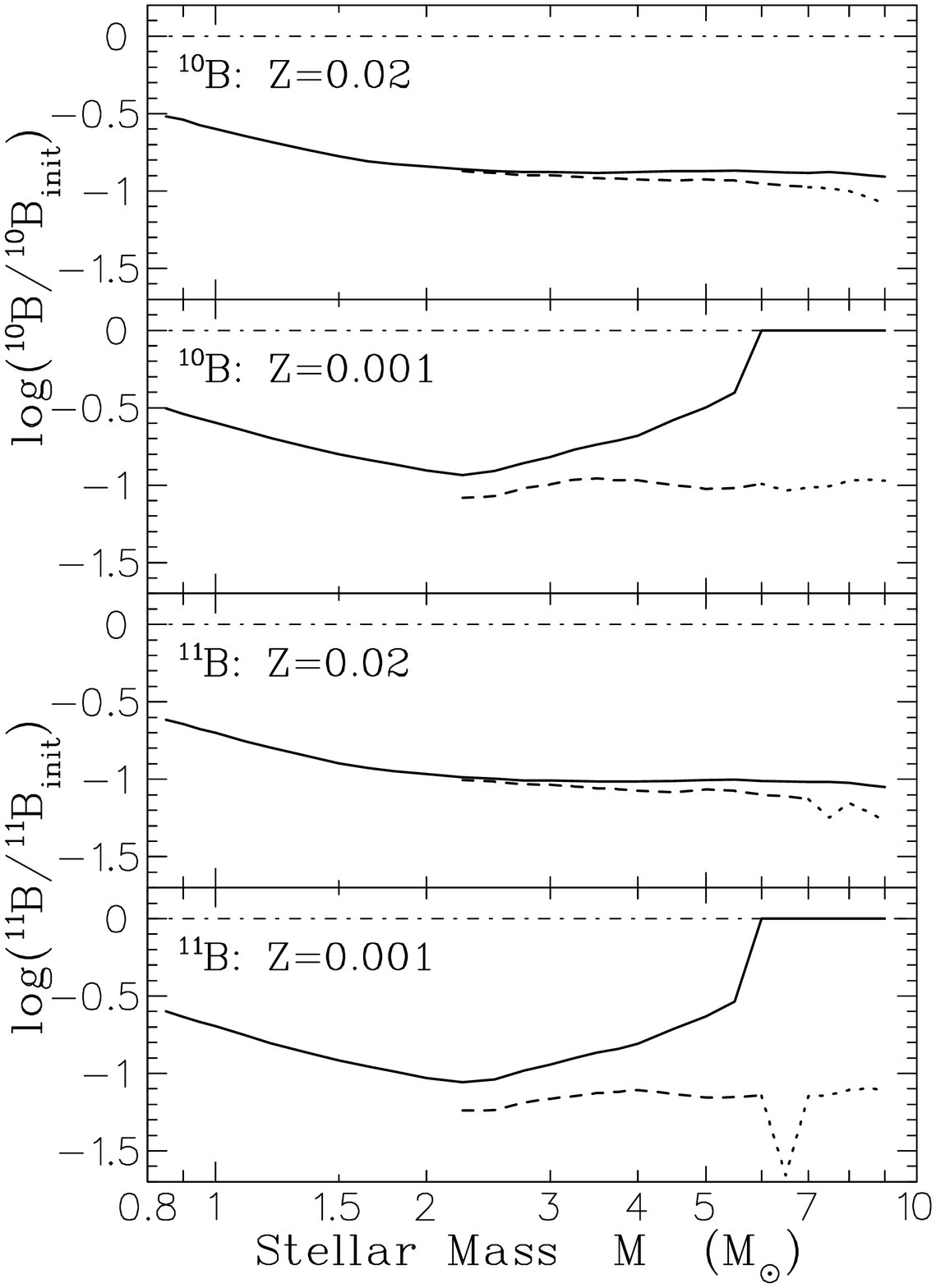}

\caption{Envelope \hbox{$\rm{}^9{Be}$}, \hbox{$\rm{}^{10}B$}, and
\hbox{$\rm{}^{11}B$} depletions due to
standard first and second dredge-up,
as a function of initial stellar mass~$M$, for solar ($Z = 0.02$) and
Population~II ($Z = 0.001$) metallicities.}

 \label{figbeb}

\end{figure}
\placefigure{figbeb}

Figure~\ref{figbeb} shows the effect of first and second dredge-up on the
surface abundances of \hbox{$\rm{}^9{Be}$}, \hbox{$\rm{}^{10}B$},
and~\hbox{$\rm{}^{11}B$}, for both solar and
Population~II compositions.  For first dredge-up, the abundance changes
are entirely due to dilution; since the extent of the pockets being
diluted is greater than for~\hbox{${}^{7}{\rm L}$}i, the dilution is
less.  For solar metallicity, this dilution is a factor
of~$\sim 30$ for~\hbox{$\rm{}^9{Be}$}, and a factor of~$\sim 10$
for \hbox{$\rm{}^{10}B$} and~\hbox{$\rm{}^{11}B$}; this dilution is
fairly independent of stellar mass except for~$M < 2\>M_\odot$ (as
was the case for~\hbox{$\rm{}^7{Li}$}).
The effect of second dredge-up is minor compared to that of
first dredge-up; there is burning
of~\hbox{$\rm{}^9{Be}$} during second dredge-up by as much as several
orders of magnitude in stars of~$\sim 7\>M_\odot$, but relatively
little burning of~\hbox{$\rm{}^{11}B$}, and no significant burning
of~\hbox{$\rm{}^{10}B$}.

There is very little data published on beryllium and boron observations
in evolved stars of low or intermediate mass.
Boesgaard, Heacox, \& Conti (1977)\markcite{BoeHC77}
searched for beryllium in four giants in the Hyades cluster, and were only
able to obtain upper limits, namely, $\rm Be/H \le 3.1 \times 10^{-13}$.
For eight Hyades dwarfs, they measured $\rm Be/H \approx 1.0 \times 10^{-11}$
(Boesgaard et al.\ 1977\markcite{BoeHC77};
Boesgaard \& Budge 1989\markcite{BoeB89}).
This implies that the giants had a beryllium dilution factor of~$\ge 32$.  From
Gilroy (1989)\markcite{Gil89},
the turn-off mass of the Hyades is about $2.2\>M_\odot$; for this mass,
our work shows that standard first dredge-up would give a dilution factor
of about~28, which is consistent.  Note that possible non-LTE effects were
not considered in the above observations.

Note that boron abundances on the RGB should be at least a factor of~$\sim 30$
higher than beryllium abundances there, due to its higher initial abundance.
If there were no reduction of the main sequence pockets from extra mixing,
there would be an additional factor of~3 from the fact that the boron pocket
is three times as big (in mass extent) as the beryllium pocket
(see Table~\ref{tblpocsiz} or Fig.~\ref{figprof}, or the first dredge-up
dilution factors from Fig.~\ref{figbeb}).  On the other hand, it is more
difficult to observe boron: the 2498$\>$\AA\ line can only be observed from
space, and in addition red giant stars have very little flux in the~UV
(compare to beryllium, where the 3130$\>$\AA\ line can be observed using
ground-based telescopes, and there is somewhat more flux).  Recently,
Duncan et al.\ (1994\markcite{Dun+94}, 1998\markcite{Dun+98})
made the first observations of boron abundances in two red giants in the
Hyades.  They find [B/H]${} = -1.1 \pm 0.2$ in the giants, as opposed to
[B/H]${} = -0.1 \pm 0.2$ on the main sequence, yielding a
boron depletion of order~10, consistent with first dredge-up predictions.
They point out that corrections for non-LTE effects for boron in giants are
yet to be calculated.


\subsection{Deep Circulation on the RGB and AGB} \label{cbp}

Observations indicate that surface \hbox{$\rm{}^{13}C$} abundances continue
to change on the RGB {\it after\/} the end of conventional first dredge-up
(Gilroy \& Brown 1991\markcite{GilB91};
Charbonnel 1994\markcite{Char94};
Charbonnel et al.\ 1998\markcite{CharBW98}),
suggestive of nuclear processing resulting from extra mixing below the base
of the convective envelope, i.e., cool bottom processing (CBP)\hbox{}.
Slow deep circulation tends to be opposed by a gradient in the mean molecular
weight~$\mu$.  A large $\mu$-discontinuity is left behind by first dredge-up.
For Population~I stars, the post-dredge-up decrease in
\hbox{$\rm{}^{12}C$}/\hbox{$\rm{}^{13}C$} is observed to commence on the RGB
only after the point where the hydrogen shell has subsequently
burned its way out to (and destroyed) this ``$\mu$-barrier'' (see, e.g.,
Charbonnel 1994\markcite{Char94};
Charbonnel et al.\ 1998\markcite{CharBW98}).
As discussed in
Boothroyd \& Sackmann (1998)\markcite{BS98},
the \hbox{$\rm{}^{12}C$}/\hbox{$\rm{}^{13}C$} observations suggest that
CBP begins when the ``$\mu$-barrier'' is erased, and then ``tails off''
in a manner intermediate between the behavior assumed in our ``single
episode'' and ``evolving~RGB'' CBP models.
Unfortunately, the data are sufficiently
sparse that this scenario is not completely certain --- for example, an
earlier starting point for CBP cannot be {\it completely\/} ruled out,
though it seems unlikely.
Note that the depth of extra mixing in our CBP models was normalized to
reproduce the observed ratio
$\hbox{$\rm{}^{12}C$}/\hbox{$\rm{}^{13}C$} \sim 13$
in the galactic open cluster~M67, which has a turn-off mass
of~$\sim 1.2\>M_\odot$ (see~\S~\ref{methods}).

The relatively sparse observations of field Population~II stars are
consistent with the
scenario in which CBP does not begin until the $\mu$-barrier is erased
(Sneden, Pilachowski, \& VandenBerg 1986\markcite{SnePV86};
Pilachowski et al.\ 1993\markcite{PilSB93}, 1997\markcite{Pil+97}),
and are also consistent
with our ``evolving RGB'' CBP models.  However, as discussed in
Boothroyd \& Sackmann (1998)\markcite{BS98},
Population~II stars in globular clusters exhibit signs of continuous
CBP throughout most or all of the RGB evolution, starting
well before the hydrogen shell has erased the $\mu$-barrier (possibly
even before deepest first dredge-up), and exhibit more
processing at a given metallicity than field Population~II stars (see, e.g.,
Carbon et al.\ 1982\markcite{Carb+82};
Trefzger et al.\ 1983\markcite{Tref+83};
Langer et al.\ 1986\markcite{Lang+86};
Suntzeff \& Smith 1991\markcite{SuntS91};
Boothroyd \& Sackmann 1998\markcite{BS98}).
Since our models made the assumption that no CBP took
place until the $\mu$-barrier was erased, they may be expected to
underestimate the amount of processing in globular cluster stars.


The ``evolving RGB'' models make the reasonable assumption that the extra
mixing process continues throughout the subsequent
RGB evolution; note that if this is the case,
then extra mixing and CBP might occur on the AGB as well.
The stellar structure on the AGB of a low mass star is similar to the
structure near the tip of the RGB, and thus lithium abundances on the AGB
might be produced at a level similar to those predicted at the tip of the
RGB by the ``evolving RGB'' model.  On the other hand, the
``single episode'' models assume that extra mixing
terminates before the star's luminosity increases much (e.g., due to
spin-down resulting from the RGB mass loss).  This is also a possible
scenario, but little or no CBP would be expected on the AGB in this case
(the driving mechanism being exhausted on the RGB\hbox{}).  The
post-RGB lithium abundance in this case would be expected to be nearly
constant, at whatever value was left behind when CBP turned off on
the RGB (probably a low lithium abundance, as the mixing would
presumably slow down before stopping, and low $\dot M_p$ values yield
low lithium abundances --- see Figs.~\ref{figrgbmpt} and~\ref{figrgbevol}c).

For stellar masses above $\sim 2.3\>M_\odot$, the RGB ends before the
hydrogen shell
reaches the $\mu$-barrier, and thus no CBP is expected on the RGB;
even if some CBP did take place,
these stars spend little time on the RGB, and processing
should be negligible.  On the early~AGB, CBP cannot be extensive,
as the hydrogen shell largely stops burning;
CBP might take place later on the thermally pulsing AGB, when the
hydrogen shell burns strongly (albeit interrupted intermittently by the
helium shell flashes).  For a fuller discussion, see
Boothroyd \& Sackmann (1998)\markcite{BS98}.

In the present work, extra mixing in both Population~I and Population~II
stars are modelled via a conveyor-belt model of deep circulation similar
to that of
WBS95\markcite{WBS95},
as discussed in~\S~\ref{methods}.
Charbonnel (1995a\markcite{Char95a},b\markcite{Char95b}) and
Denissenkov \& Weiss (1996)\markcite{DenW96}
have modelled such mixing in Population~II stars using a diffusion algorithm.
As discussed in
WBS95\markcite{WBS95},
these two approaches should yield similar results for the CNO isotopes, since
the nuclear processing of CNO isotopes should be insensitive to the
speed or geometry of mixing.
However, as discussed below, the lighter elements are more
dependent on the mixing speed, especially~\hbox{$\rm{}^7{Li}$}.  They are
discussed in detail in the following sections.


\subsubsection{\hbox{$\rm{}^7{Li}$}} \label{cbpli}


\begin{figure}[!t]
  \plotfiddle{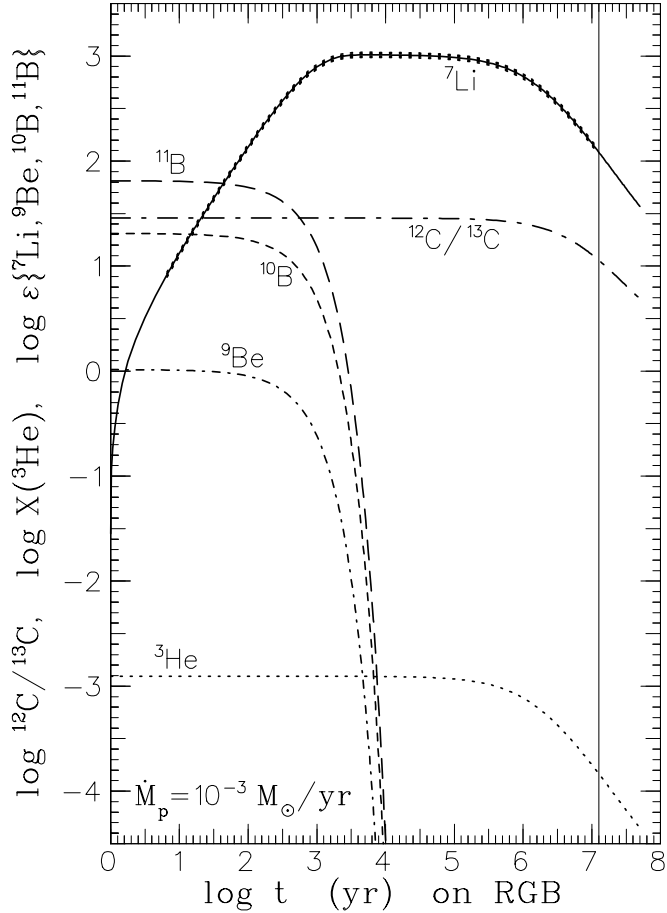}{4.5 true in}{0}{52}{52}{-180}{-43}

\caption{The effect of CBP in our ``single episode'' models,
as a function of time on the RGB\hbox{}.
These are for a solar metallicity $1\>M_\odot$ star with an envelope mass
$M_{\rm env} = 0.7\>M_\odot$ and a mixing rate of deep circulation
$\dot M_p = 10^{-3}\>M_\odot$/yr; circulation was assumed to reach to
$\Delta \log\,T = 0.17$ from the base of the hydrogen-burning shell,
such that the envelope \hbox{$\rm{}^{12}C/{}^{13}C$}
ratio reaches the observed value of $\sim 11$ after a time
$t_{\rm mix:RGB} \sim 1.25 \times 10^7\>$yr ({\it vertical line\/}),
although the computations were continued for $5 \times 10^7\>$yr.
The quantities plotted are the log of the
\hbox{$\rm{}^{12}C$}/\hbox{$\rm{}^{13}C$} number ratio,
the log of the \hbox{$\rm{}^3{He}$} mass fraction, and
$\log\,\varepsilon$ values for the other light elements.
{\it Hatched\/} regions of \hbox{$\rm{}^7{Li}$} curves emphasize cases where
abundances are higher than the upper bound of typical RGB field stars.}

 \label{figrgbt}

\end{figure}
\placefigure{figrgbt}


\begin{figure}[!t]
  \plotfiddle{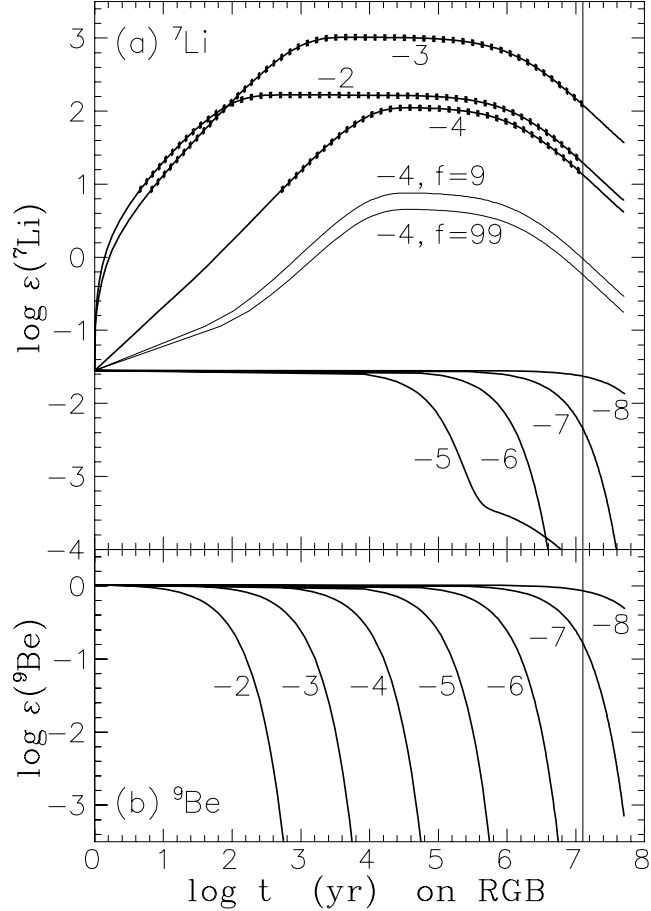}{4.5 true in}{0}{52}{52}{-180}{-43}

\caption{(a)~Dependence of the \hbox{$\rm{}^7{Li}$} creation and depletion
on the mixing speed $\dot M_p$ and on our geometry factor $f \equiv f_u/f_d$
in our ``single episode'' CBP models on the RGB\hbox{}.  Curves are
labelled by $\log\,\dot M_p$, and by~$f$ for cases with $f \ne 1$; hatched
regions of curves emphasize high lithium abundance cases (as in
Fig.~\protect\ref{figrgbt}).  Note that models with
$\dot M_p < 10^{-7}\>M_\odot$/yr do not produce enough
\hbox{$\rm{}^{13}C$} to match the observations by the time~$t_{\rm mix:RGB}$
is reached (indicated by {\it vertical line\/}).
(b)~Dependence of \hbox{$\rm{}^9{Be}$} destruction,
similarly.  Note that boron isotope destruction curves would have the same
form, except for having higher initial abundance.}

 \label{figrgbmpt}

\end{figure}
\placefigure{figrgbmpt}


\begin{figure}[!t]
     \bootplotfidtwo{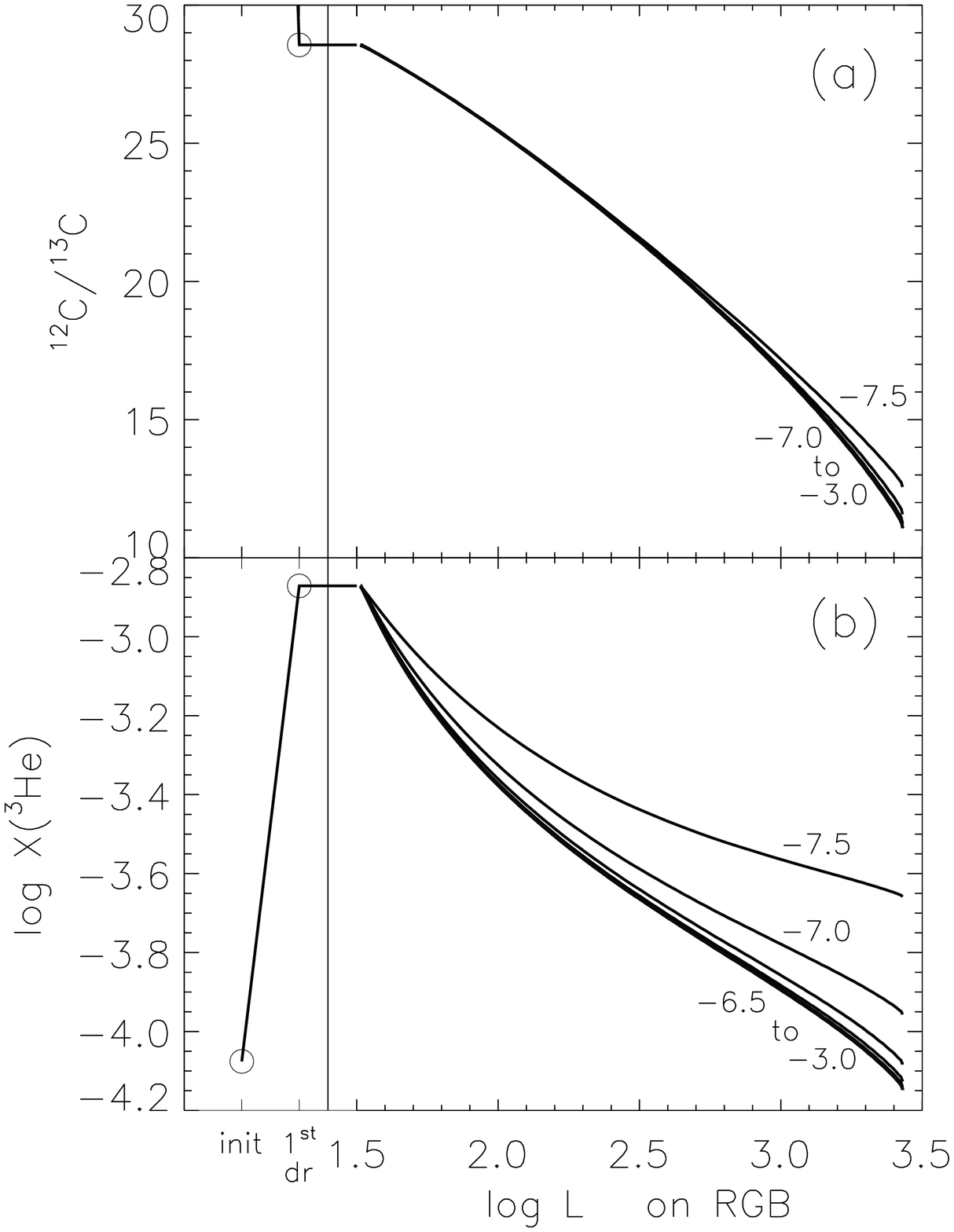}{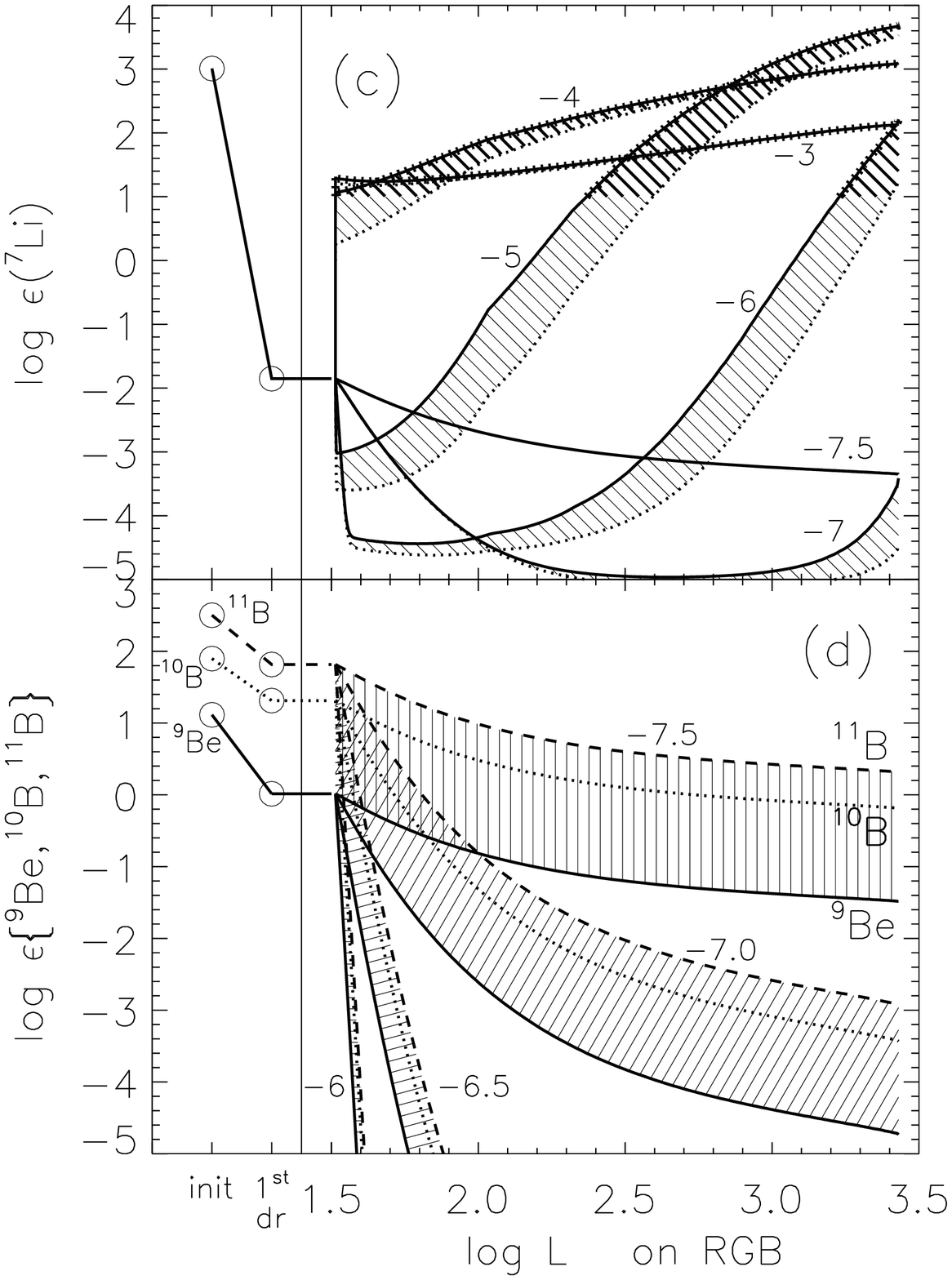}{4 true
       in}{48}{-180}{-40}{48}{-105}{-40}

\caption{The effect of CBP (for various mixing
speeds~$\dot M_p$) in our ``evolving RGB'' solar metallicity $1\>M_\odot$
CBP models as a function of luminosity on the RGB; curves are labelled by
$\log\,\dot M_p$.  Initial and first dredge-up abundances are shown
schematically in the left portion of each panel (open circles).
(a)~The \hbox{$\rm{}^{12}C/{}^{13}C$} number ratio; the initial ratio
(offscale) was~90.
(b)~The log of the \hbox{$\rm{}^{3}He$} mass fraction.
(c)~The \hbox{$\log\,\varepsilon({}^7{\rm Li})$} values.  The filling
between cases with geometry factors $f = 1$ ({\it solid curves}) and $f = 9$
({\it dotted curves}) having the same $\dot M_p$ indicates the effect of
varying the circulation geometry (see text).  Hatched
regions of curves and heavy filled regions emphasize high lithium abundance
cases, as in Fig.~\protect\ref{figrgbt}.
(d)~Values of \hbox{$\log\,\varepsilon({}^9{\rm Be})$} ({\it solid curves}),
\hbox{$\log\,\varepsilon({}^{10}{\rm B})$} ({\it dotted curves}), and
\hbox{$\log\,\varepsilon({}^{11}{\rm B})$} ({\it dashed curves});
low-$\dot M_p$ cases are omitted for clarity.  Curves for Be and B having
the same $\dot M_p$ value are connected by filled regions.}

 \label{figrgbevol}

\end{figure}
\placefigure{figrgbevol}

Figure~\ref{figrgbt} shows the abundances of the light elements (and
\hbox{$\rm{}^{12}C/{}^{13}C$}) as a function of time on the RGB, for one of
our ``single episode'' CBP models (note that
\hbox{$\rm{}^{12}C/{}^{13}C$} reaches the observed value after
$1.25 \times 10^7\>$yr, but most of the activity in the light elements takes
place much earlier; computations were terminated after $5 \times 10^7\>$yr).
It is a key diagram, demonstrating that under certain conditions
a major amount of \hbox{$\rm{}^7{Li}$} can be created on the RGB due to
CBP in a $1\>M_\odot$ star of solar metallicity.
Figure~\ref{figrgbmpt}a shows the \hbox{$\rm{}^7{Li}$} abundances similarly,
for various mixing speeds~$\dot M_p$.  One sees that \hbox{$\rm{}^7{Li}$} is
enhanced for models with rapid enough circulation (namely,
$\dot M_p \gtrsim 10^{-4}\>M_\odot$/yr); lower mixing speeds yield
lithium destruction.

Figure~\ref{figrgbevol} shows the abundances of the light elements (and
\hbox{$\rm{}^{12}C/{}^{13}C$}) as a function of luminosity on the RGB, for
our ``evolving RGB'' CBP models with various mixing
speeds~$\dot M_p$.  Note that in these models the total mass drops from 0.984
to~$0.710\>M_\odot$ (due to mass loss on the RGB), the core mass grows from
0.243 to~$0.462\>M_\odot$, and the convective envelope mass~$M_{\rm env}$ thus
drops from 0.718 to~$0.246\>M_\odot$ (for lower metallicities or higher
stellar masses, these changes are smaller).
In these models, the value reached by \hbox{$\rm{}^{12}C/{}^{13}C$}
at the tip of the RGB agrees with the observed values (i.e., the circulation
was assumed to operate throughout the~RGB); it is essentially independent of
the mixing speed or geometry (see Fig.~\ref{figrgbevol}a).  As in the
``single episode'' cases, the \hbox{$\rm{}^7{Li}$} abundance
is rapidly enhanced for $\dot M_p \gtrsim 10^{-4}\>M_\odot$/yr; but even at
lower mixing speeds ($\dot M_p \gtrsim 10^{-6}\>M_\odot$/yr), significant
\hbox{$\rm{}^7{Li}$} enhancements occur as the star climbs the~RGB\hbox{}.

The present work shows for the first time that {\it low mass RGB stars\/}
can also become super-rich lithium stars.  This allows one to understand
the ``K~Giant Lithium Problem''
(de~la~Reza \& da~Silva 1995\markcite{delaRS95}),
of anomalously high lithium abundances
discovered in otherwise normal low-mass K~giants of relatively low luminosity
(Wallerstein \& Sneden 1982\markcite{WalS82};
Hanni 1984\markcite{Han84};
Brown et al.\ 1989\markcite{Bro+89};
Gratton \& D'Antona 1989\markcite{GraD89};
Pilachowski et al.\ 1990\markcite{PilSH90};
Pallavicini et al.\ 1990\markcite{Pal+90};
Fekel \& Marschall 1991\markcite{FekM91};
Fekel \& Balachandran 1993\markcite{FekB93};
da~Silva et al.\ 1995a\markcite{daSRB95a},b\markcite{daSRB95b};
de~la~Reza \& da~Silva 1995\markcite{delaRS95};
de~la~Reza et al.\ 1996\markcite{delaRDS96}, 1997\markcite{delaR+97};
Fekel et al.\ 1996\markcite{Fek+96}).
The presence and amount of
lithium creation are critically dependent on the assumed speed~$\dot M_p$
of extra mixing, as well as on the geometry of mixing
(as measured by our factor $f \equiv f_u/f_d$, the fractional area of upward
streams relative to downward streams in our conveyor-belt circulation
model).  Deeper extra mixing yields higher lithium enhancements
for a given (rapid) mixing rate (as may be seen by comparing
peak \hbox{$\rm{}^7{Li}$} values from Fig.~\ref{figrgbmpt} with the values
attained in Fig.~\ref{figrgbevol}c near $\log\,L = 1.5$).
However, it is noteworthy that the amount of \hbox{$\rm{}^7{Li}$}
created does not depend on the previous lithium abundance, i.e., on the
lithium history of the star.  A qualitative understanding of this
mechanism for RGB lithium enrichment is relatively straightforward.

In order to produce the observed
additional~\hbox{$\rm{}^{13}C$}, the extra mixing must reach temperatures
high enough that \hbox{$\rm{}^3{He}$} is also burned (as shown
in Figs.~\ref{figrgbt} and~\ref{figrgbevol}), resulting in
\hbox{$\rm{}^7{Be}$} creation
via $\hbox{$\rm{}^3{He}$}(\alpha,\gamma)\hbox{$\rm{}^7{Be}$}$.
If the extra mixing is slow, this \hbox{$\rm{}^7{Be}$} is
destroyed while still at high temperatures via
$\hbox{$\rm{}^7{Be}$}(p,\gamma)\hbox{${}^8{\rm B}$}
(\hbox{$e^{\hbox{$\scriptscriptstyle +$}}$}\nu)\hbox{$\rm{}^8{Be}$}
\rightarrow 2\alpha$
or $\hbox{$\rm{}^7{Be}$}(\hbox{$e^{\hbox{$\scriptscriptstyle -$}}$},\nu)
\hbox{$\rm{}^7{Li}$}$, and any \hbox{$\rm{}^7{Li}$} produced from
\hbox{$\rm{}^7{Be}$} electron capture is immediately burned up via
$\hbox{$\rm{}^7{Li}$}(p,\alpha)\hbox{$\rm{}^4{He}$}$.  However, for
sufficiently high mixing speeds,
\hbox{$\rm{}^7{Be}$}~can be transported out to cooler regions before the
electron capture takes place, where the resulting \hbox{$\rm{}^7{Li}$}
can survive and enrich
the stellar envelope.  In other words, {\it under special conditions
the Cameron-Fowler mechanism can work in low-mass, low-luminosity
RGB stars}.  The resulting envelope \hbox{$\rm{}^7{Li}$} abundance is
determined by the balance between \hbox{$\rm{}^7{Be}$} being transported
out (to form~\hbox{$\rm{}^7{Li}$}), and \hbox{$\rm{}^7{Li}$} being
transported back inwards (and being destroyed).  The envelope
\hbox{$\rm{}^7{Li}$} abundance reaches equilibrium relative to the abundance
of~\hbox{$\rm{}^3{He}$} on a timescale $t_{proc} \sim M_{\rm env} / \dot M_p$,
the time required for circulation to process the entire envelope; $t_{proc}$~is
the timescale on which the \hbox{$\rm{}^7{Li}$} abundance reaches its peak
values in Figures~\ref{figrgbt}, \ref{figrgbmpt}a, and~\ref{figrgbevol}c.

The effect on the envelope \hbox{$\rm{}^7{Li}$} abundance of changing the
mixing speed~$\dot M_p$ or the relative areas of upward and downward
streams~$f \equiv f_u/f_d$ can be readily understood.
For mixing speeds $\dot M_p > 10^{-7}\>M_\odot$/yr, only a small fraction
of the \hbox{$\rm{}^3{He}$} in a circulating blob of matter is destroyed during
a single circulation pass, and the total \hbox{$\rm{}^3{He}$} burning rate
per unit time is then independent of~$\dot M_p$ (see
Fig.~\ref{figrgbevol}b): the amount burned per
circulation pass is then inversely proportional to~$\dot M_p$, and the
number of circulation passes per unit time is proportional to~$\dot M_p$.
The amount of \hbox{$\rm{}^7{Be}$}
produced per unit time is thus independent of~$\dot M_p$, but the amount that
survives to reach cool temperatures is not.  If one increases~$\dot M_p$,
at first the amount of surviving \hbox{$\rm{}^7{Be}$} increases, as does
the resulting \hbox{$\rm{}^7{Li}$} production.  However, as $\dot M_p$
increases still more, there comes a point where essentially all the
\hbox{$\rm{}^7{Be}$} survives; the \hbox{$\rm{}^7{Li}$} production rate
cannot increase further.  The \hbox{$\rm{}^7{Li}$} destruction rate,
however, still increases with increasing~$\dot M_p$ (since all
\hbox{$\rm{}^7{Li}$} transported down is burned), and the resulting
envelope \hbox{$\rm{}^7{Li}$} abundance decreases with increasing~$\dot M_p$.
The value of $\dot M_p^m$ which yields the maximum \hbox{$\rm{}^7{Li}$}
production depends on the temperature at the base of the circulation (as may
be seen from Fig.~\ref{figrgbevol}c, where this temperature grows as the
star ascends the RGB\hbox{}).  If the circulation speed is less
than~$\dot M_p^m$, then changing the stream geometry will affect
the~\hbox{$\rm{}^7{Li}$} abundance.  Increasing~$f_u/f_d$ yields a wider and
slower upward stream; less of the \hbox{$\rm{}^7{Be}$} produced at the base
of the circulation survives out to cool regions, yielding a lower
\hbox{$\rm{}^7{Li}$} abundance (see Fig.~\ref{figrgbevol}c).

Figures~\ref{figrgbt} and~\ref{figrgbevol} show that the peak
\hbox{$\rm{}^7{Li}$} abundance can be attained prior to any significant
\hbox{$\rm{}^{13}C$} enrichment, but that the \hbox{$\rm{}^7{Li}$} abundance
can remain high or even grow during
the \hbox{$\rm{}^{13}C$} enrichment process.  These preliminary models
would thus predict that low mass super-rich lithium stars could occur
with \hbox{$\rm{}^{12}C/{}^{13}C$} ratios throughout the
range from $\sim 30$ to~$\sim 4$.  This is in reasonable agreement with
the observations presented by
da~Silva et al.\ (1995a\markcite{daSRB95a},b\markcite{daSRB95b}),
and with the statement of
de~la~Reza et al.\ (1996)\markcite{delaRDS96}
that ``the richest Li stars show the smallest
\hbox{$\rm{}^{12}C$}/\hbox{$\rm{}^{13}C$} ratios''.
The fact that lithium-rich low-mass RGB stars are observed to be rare implies
either that few stars attain rapid enough circulation speeds for RGB lithium
production, or that rapid circulation speeds are attained only briefly.
Recent observations suggest that enhanced RGB lithium abundances occur
in conjunction with episodic mass loss in low mass stars,
for periods lasting~$\sim 10^5\>$yr
(de~la~Reza 1995\markcite{delaR95};
de~la~Reza et al.\ 1996\markcite{delaRDS96}, 1997\markcite{delaR+97};
Fekel et al.\ 1996\markcite{Fek+96};
see also
Wallerstein \& Morell 1994\markcite{WalM94}).
This suggests that the extra mixing may be episodic
in nature, rather than continuous,
providing an additional complication.  Short episodes of mixing to a depth
rather deeper than assumed in our models could yield
\hbox{$\rm{}^{12}C/{}^{13}C$} ratios in agreement with the observations,
\hbox{$\rm{}^3{He}$}~depletion to an extent comparable to that in our models,
and \hbox{$\rm{}^7{Li}$} abundances that could be even larger than in our
models (depending on the speed and depth of mixing).  Most of the
\hbox{$\rm{}^7{Li}$} produced this way would be subsequently destroyed, as
the mixing episode died away.
An alternative explanation of the above observations invokes variations in
the speed (and possibly the depth) of extra mixing, without requiring that
it die away completely.

The amount of \hbox{$\rm{}^7{Li}$} injected into the interstellar medium
by the observed lithium-rich K~giants depends strongly on the parameters
of the lithium enrichment scenario, namely, the size and timescales of the
lithium enhancements, the magnitude and timescales of the mass loss, and
whether, how often, and at what points on the RGB such enhancement episodes
recur.  For typical parameters from the scenario of
de~la~Reza et al.\ (1996\markcite{delaRDS96}, 1997\markcite{delaR+97}),
(namely, $\log\,\varepsilon({}^7{\rm Li}) \lesssim 4$ for timescales
$\lesssim 10^5\>$yr, mass loss of $\sim 10^{-7}\>M_\odot$/yr for timescales
of $\sim 200\>$yr and $\lesssim 10^{-10}\>M_\odot$/yr thereafter, and
recurrence of $\sim 10$ times per star at $\log\,L \sim 2$), the
{\it average\/} \hbox{$\rm{}^7{Li}$} abundance of the total amount
of material ejected from low mass stars would be less than 1\% of the cosmic
abundance --- a negligible amount.  However, if at least some
\hbox{$\rm{}^7{Li}$} enrichment episodes occurred near the tip of the RGB,
this average \hbox{$\rm{}^7{Li}$} abundance could be more than an
order of magnitude larger,
due to the high mass loss rates ($\sim 10^{-7}\>M_\odot$/yr) maintained
at the RGB tip for the entire remaining RGB time ($\sim 10^5$ -- $10^6\>$yr),
and could yield non-negligible \hbox{$\rm{}^7{Li}$} enrichment of the
interstellar medium.
In other words, \hbox{$\rm{}^7{Li}$} production via CBP could possibly
have a significant impact on the interstellar medium; further observations
are needed to pin down the RGB \hbox{$\rm{}^7{Li}$} enhancements
simultaneously with the mass loss.


\begin{figure}[!t]
  \plotfiddle{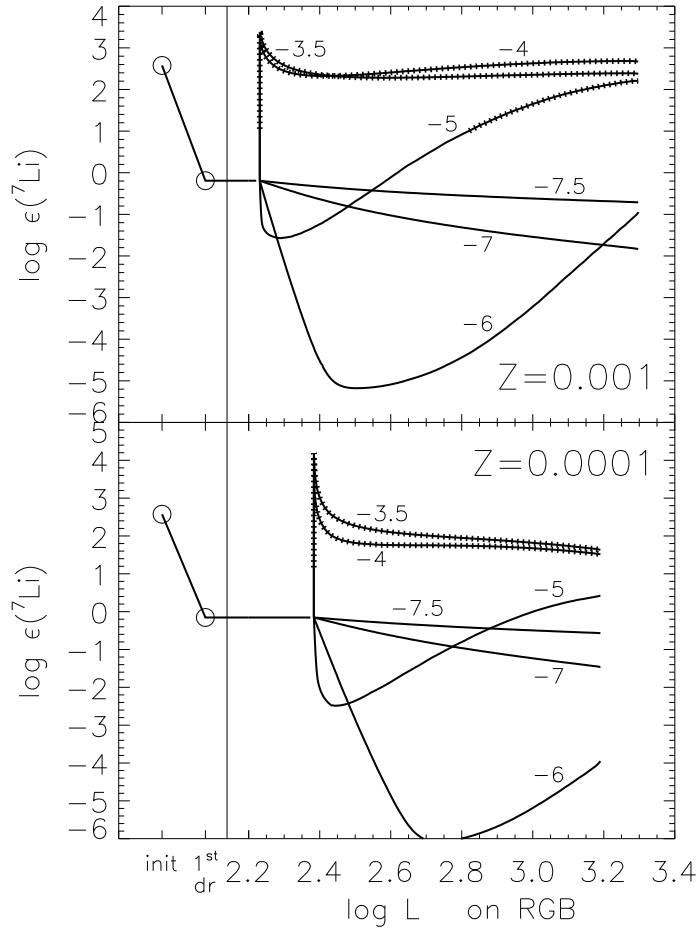}{4.5 true in}{0}{54}{54}{-180}{-43}

\caption{The effect of CBP on \hbox{$\rm{}^7{Li}$} in our ``evolving RGB''
Population~II models of~$1\>M_\odot$; curves are labelled by the mixing
speed, $\log\,\dot M_p$.  Initial and first dredge-up abundances are
shown schematically at left,
as in Fig.~\protect\ref{figrgbevol}; hatched regions of curves emphasize high
lithium abundance cases, as in Fig.~\protect\ref{figrgbt}.}

 \label{figcbploz}

\end{figure}
\placefigure{figcbploz}

Note that Population~II stars are expected to experience more vigorous
CBP than Population~I stars, with the extra mixing
reaching higher temperatures
(Boothroyd \& Sackmann 1998\markcite{BS98}).
Figure~\ref{figcbploz} shows that low-mass Population~II red giants
can also become super-rich lithium stars.  For high mixing rates
($\dot M_p \gtrsim 10^{-4}\>M_\odot$/yr),
\hbox{$\log\,\varepsilon({}^7{\rm Li})$}${} \sim 4$ can be attained,
a considerably higher abundance than the star was endowed with at birth
(\hbox{$\log\,\varepsilon({}^7{\rm Li})$}${} \sim 2$--3) and three orders
of magnitude higher than the value
\hbox{$\log\,\varepsilon({}^7{\rm Li})$}${} \sim 1$ observed in typical
Population~II subgiants
(Pilachowski et al.\ 1993\markcite{PilSB93}).
Such extremely high lithium abundances can exist only at the onset of CBP,
since the \hbox{$\rm{}^3{He}$} fuel for lithium creation is rapidly
depleted; however, lithium abundances high enough to be detectable can exist
all the way up the RGB (see Fig.~\ref{figcbploz}).  For intermediate
mixing rates ($\dot M_p \sim 10^{-5}\>M_\odot$/yr), lithium is destroyed
at the onset of CBP, but near the tip of the RGB enough
may be created to be detectable.  For low mixing rates
($\dot M_p \sim 10^{-6}\>M_\odot$/yr), lithium is simply destroyed;
for very low mixing rates ($\dot M_p \sim 10^{-7}\>M_\odot$/yr), lithium
destruction is slow enough that lithium may remain detectable for much
or even all of the RGB\hbox{}.  The effect of varying the geometry factor~$f$
was not calculated explicitly in these models, but would be similar to the
effect shown (for solar metallicity) in Figure~\ref{figrgbevol}c.  The trends
of Figure~\ref{figcbploz}, relative to Figure~\ref{figrgbevol}c, would be
exaggerated in any stars where extra mixing reached higher temperatures than
assumed in our models; there is observational evidence that this is the case
in some Population~II stars in globular clusters, as discussed above (see also
Boothroyd \& Sackmann 1998\markcite{BS98}).

Note that the observations of field Population~II subgiants and giants by
Pilachowski et al.\ (1993)\markcite{PilSB93}
find additional lithium depletion on the RGB subsequent to first dredge-up,
at the effective temperature where the ``$\mu$-barrier'' has been erased
and CBP is expected to start
(Charbonnel 1994\markcite{Char94}, 1995a\markcite{Char95a},b\markcite{Char95b};
Charbonnel et al.\ 1998\markcite{CharBW98};
Boothroyd \& Sackmann 1998\markcite{BS98}).
The lithium depletion in these stars was reproduced by
Charbonnel's (1995a\markcite{Char95a},b\markcite{Char95b})
interesting deep mixing models (which we therefore infer to have had low
mixing speeds); they can also be reproduced by those Population~II
``evolving RGB'' models of the present work that have $\dot M_p \lesssim
10^{-5}\>M_\odot$/yr.  Note that
Pilachowski et al.\ (1993)\markcite{PilSB93}
observed 15 Population~II RGB stars showing extra lithium depletion
(beyond that from first dredge-up), but no stars showing lithium
enhancement.  It appears that lithium enhancement in field Population~II red
giants is a rare phenomenon.

If CBP also occurred on the AGB
(BSW95\markcite{BSW95};
WBS95\markcite{WBS95};
Boothroyd \& Sackmann 1998\markcite{BS98}),
any remaining \hbox{$\rm{}^3{He}$} might then be converted into
\hbox{$\rm{}^7{Li}$}, resulting in lithium-rich low-mass AGB stars.
Low mass extreme Population~II stars, which destroy all their
\hbox{$\rm{}^3{He}$} on the RGB (see $Z = 0.0001$ case in Fig.~\ref{fighe3}),
would never be expected ever to become lithium-rich on the~AGB\hbox{}.



\subsubsection{\hbox{$\rm{}^9{Be}$}, \hbox{$\rm{}^{10}B$},
 and \hbox{$\rm{}^{11}B$}} \label{cbpbeb}

Unlike \hbox{$\rm{}^7{Li}$}, the isotopes \hbox{$\rm{}^9{Be}$},
\hbox{$\rm{}^{10}B$}, and~\hbox{$\rm{}^{11}B$} cannot be created
by CBP; they can only be destroyed.  For mixing rates rapid enough
that the entire envelope can be processed on the RGB (i.e.,
$t_{proc} \sim M_{\rm env} / \dot M_p \ll \tau_{\scriptscriptstyle\!\rm RGB}$,
or $\dot M_p \gtrsim 10^{-7}\>M_\odot$/yr), these isotopes are completely
destroyed (see Figs.~\ref{figrgbt}, \ref{figrgbmpt}b, and~\ref{figrgbevol}d);
they can be partially preserved
only if the mixing is slow enough that only part of the envelope is processed.
As may be seen in Figures~\ref{figrgbt} and~\ref{figrgbevol}, the boron
isotopes behave very much like~\hbox{$\rm{}^9{Be}$}.  Figure~\ref{figrgbmpt}
shows that there exists only a very brief interval when high
lithium abundances co-exist with non-zero beryllium (and boron) abundances,
as one would expect, since the former is created on the same
timescale~$t_{proc}$ that the latter are destroyed.
Thus low mass lithium-rich RGB stars should almost without exception be
completely depleted in beryllium and boron, provided that CBP
operates continuously as assumed by our models.  If the
CBP consists of relatively short-lived episodes
(possibly repeated), with each mixing episode reaching higher temperatures
than required by a continuous mixing model, then the
observed \hbox{$\rm{}^{13}C$} enhancements in Population~I stars
could be produced in conjunction with enhanced lithium
abundances, but large depletions of beryllium and boron need not occur (as
only part of the envelope need be mixed down in each episode, due to the
short timescale).  Thus if one observes enhanced lithium combined with
non-zero beryllium and boron, this is probably a signature of episodic
deep mixing.  So far, no beryllium or boron observations have been made
for stars in this CBP stage on the RGB\hbox{}.


\subsubsection{\hbox{$\rm{}^3{He}$}}

Figures~\ref{figrgbt} and~\ref{figrgbevol} show that, provided the CBP
is strong enough to create the observed enhancement of
\hbox{$\rm{}^{13}C$} in a $1\>M_\odot$ star of solar metallicity, there must
be an accompanying partial destruction of~\hbox{$\rm{}^3{He}$}.  This
\hbox{$\rm{}^3{He}$} destruction is by a factor of order~10 at the
appropriate time ($t \approx 1.25 \times 10^7\>$yr) in
our ``single episode''
CBP models, and by a factor of order~20 in our ``evolving
RGB'' CBP models, for most of the possible range of the speed of extra mixing
(namely, $\dot M_p \gtrsim 10^{-7}\>M_\odot$/yr).  Note that
\hbox{$\rm{}^{12}C/{}^{13}C$} observations as a function of RGB luminosity
in low mass stars suggest a scenario somewhere between these two models,
as discussed in
Boothroyd \& Sackmann (1998)\markcite{BS98}.
Since first dredge-up causes a \hbox{$\rm{}^3{He}$} enrichment by a
factor of~$\sim 15$ in a $1\>M_\odot$ star (see Fig.~\ref{fighe3}),
CBP results in a reduction of the \hbox{$\rm{}^3{He}$} abundance to roughly
its initial value.  There is less depletion at higher stellar masses, but
more depletion at lower metallicities, as shown in Figure~\ref{fighe3}.
At a minimum, the observed \hbox{$\rm{}^{13}C$} enhancements in solar
metallicity stars imply significant \hbox{$\rm{}^3{He}$} depletion;
low-metallicity stars, with their large observed carbon depletions on the
RGB, must also be almost completely depleted in~\hbox{$\rm{}^3{He}$}.
Thus low-metallicity stars must be net destroyers of~\hbox{$\rm{}^3{He}$}
when the contributions of all stellar masses are included,
rather than net producers, as standard first
dredge-up theory would predict.  Solar metallicity stars may be net destroyers
of~\hbox{$\rm{}^3{He}$}, and surely do not produce very much (again, in
contrast to the predictions of standard first dredge-up theory).
Note that destruction of \hbox{$\rm{}^3{He}$} has also been obtained by
Charbonnel (1995a\markcite{Char95a},b\markcite{Char95b})
in models of extra mixing in low mass metal-poor stars.

One can make more quantitative predictions.  First, consider the predictions
of standard first and second dredge-up from Figure~\ref{fighe3}; note also
that hot bottom burning yields almost complete \hbox{$\rm{}^3{He}$}
destruction in intermediate mass stars ($4 - 7\>M_\odot$ for Population~I
stars, and $3.5 - 6\>M_\odot$ for Population~II stars
[Boothroyd \& Sackmann 1992\markcite{BS92}]).
Weaver \& Woosley (1993)\markcite{WeaW93}
estimate that supernovae
(from stars with mass $\sim 12 - 40\>M_\odot$) eject material that is
depleted in \hbox{$\rm{}^3{He}$} by a factor between 2 and~4.  Weighting
by the fraction of each star's mass that is ejected and by a
Salpeter (1955)\markcite{Sal55}
initial mass function (IMF) with a typical exponent of~$s \approx 2.3 \pm 0.2$,
implies that standard first and second dredge-up would result in overall
\hbox{$\rm{}^3{He}$} enhancement by a factor $g_3^{\rm dr} \sim 2.4 \pm 0.5$
in Population~I stars with masses between 1 and~$40\>M_\odot$.
If one considers as well the \hbox{$\rm{}^3{He}$} destruction from our
``evolving RGB'' CBP models, one obtains an overall stellar
\hbox{$\rm{}^3{He}$} survival fraction $g_3 \approx 0.8 - 0.9$
(depending on the depletion in supernovae), while our ``single episode'' CBP
models would yield $g_3 \approx 1.0 - 1.1$; an uncertainty of $\pm 0.2$ in the
Salpeter IMF exponent~$s$ yields an additional uncertainty of only
$\pm 0.1$ in~$g_3$.  Population~II stars, with stronger CBP, would have
lower \hbox{$\rm{}^3{He}$} survival fractions: $g_3 \sim 0.6$ for $Z = 0.001$,
and $g_3 \sim 0.35$ for $Z = 0.0001$.

A star of $1\>M_\odot$ loses a large fraction of its envelope mass while still
on the~RGB\hbox{}.  Stars of higher mass, however, retain most of their mass
until the AGB; thus one may consider the possibility that CBP occurs on
the AGB as well as the RGB.  Including CBP on the AGB
in solar metallicity stars with the same parameters as the ``evolving RGB''
models yields $g_3 \approx 0.7 - 0.8$; Population~II stars would have
$g_3 \sim 0.55$ for $Z = 0.001$, and $g_3 \sim 0.3$ for $Z = 0.0001$.
If one assumed instead that CBP on the AGB
should reproduce observed \hbox{$\rm{}^{18}O$} depletions in low mass AGB
stars (which, however, seems unlikely: see
Boothroyd \& Sackmann [1998]\markcite{BS98}),
then AGB cool bottom processing would have to
be much stronger, yielding $g_3 \sim 0.3$ for solar metallicity
and $g_3 \sim 0.2$ for Population~II stars.

Recent measurements of high \hbox{$\rm{}^3{He}$} abundances in a few
planetary nebulae ejected from low mass stars
(Galli et al.\ 1997\markcite{Gal+97})
indicate that not quite all low mass stars can experience CBP and
\hbox{$\rm{}^3{He}$} depletion; this could increase somewhat the above
values of~$g_3$.  Depending on the fraction of low mass RGB stars that
excape CBP and the exact extent of \hbox{$\rm{}^3{He}$} depletion in the
other RGB stars, consistent galactic chemical evolution models can
accommodate a fairly wide range of possible primordial deuterium abundances
(see, e.g.,
Olive et al.\ 1997\markcite{Oli+97}),
i.e., some relaxation is allowed of the strongest
lower limit on the cosmic baryon density~$\Omega_b$ from
Big Bang nucleosynthesis calculations.  However,
Tosi et al.\ (1998)\markcite{Tosi+98}
point out that it is very difficult to contrive a consistent galactic
chemical evolution model that depletes primordial deuterium by more than
a factor of~5, let alone the factor of~10 that would be required to
accommodate the ``high'' deuterium observations in quasars.

\section{Conclusions}

1. It has been demonstrated that \hbox{$\rm{}^7{Li}$}
{\it can be created in low mass red giant stars}, via
extra deep mixing and the associated ``{\it cool bottom processing\/}''
(namely, via the same process
which is usually invoked on the red giant branch [RGB] to explain
the observed anomalous \hbox{$\rm{}^{13}C$} enhancement,
beyond that resulting from first dredge-up).  This \hbox{$\rm{}^7{Li}$}
production can account for the recent discovery of surprisingly high
lithium abundances in some low mass red giants.

2. Lithium is created via the
Cameron-Fowler mechanism, in which \hbox{$\rm{}^7{Be}$}
created via $\hbox{$\rm{}^3{He}$}(\alpha,\gamma)\hbox{$\rm{}^7{Be}$}$ is
transported out fast enough that its electron-capture to~\hbox{$\rm{}^7{Li}$}
can take place in cool regions, where the resulting \hbox{$\rm{}^7{Li}$}
can survive to enrich the stellar envelope.  The amount
of \hbox{$\rm{}^7{Li}$} produced can exceed
$\log\,\varepsilon({}^7{\rm Li}) \sim 4$, but
depends critically on the details of
the extra mixing mechanism (mixing speeds, geometry, episodicity); it is
{\it independent\/} of the previous \hbox{$\rm{}^7{Li}$} history of the star.

3. If the deep circulation is a long-lived, continuous process, lithium-rich
RGB stars should be completely devoid of beryllium and boron.
If it occurs in short-lived episodes, higher \hbox{$\rm{}^7{Li}$} abundances
might result, and beryllium and boron might be only partially destroyed.

4. Under some circumstances, cool bottom
processing in low mass stars can produce {\it super-rich\/} lithium stars
(i.e., \hbox{$\rm{}^7{Li}$} abundances larger than the interstellar medium
value), analogous to hot bottom burning in intermediate mass stars.
For the interstellar medium, low mass stars might possibly be a significant
source of~\hbox{$\rm{}^7{Li}$}.

5. Cool bottom processing leads to \hbox{$\rm{}^3{He}$} destruction in low
mass stars; in contrast to the \hbox{$\rm{}^7{Li}$} creation, the net
\hbox{$\rm{}^3{He}$} depletion is largely
independent of the details of the extra mixing mechanism.
The overall contribution of solar-metallicity stars from 1 to $40\>M_\odot$
is expected to be net destruction of~\hbox{$\rm{}^3{He}$}, with an
overall \hbox{$\rm{}^3{He}$} survival fraction $g_3 \approx 0.9 \pm 0.2$
(weighted average over all stellar masses); this is in contrast to
standard dredge-up theory, which would predict that stars are net producers
of~\hbox{$\rm{}^3{He}$} (with~$g_3^{\rm dr} \sim 2.4 \pm 0.5$).  However,
high measured \hbox{$\rm{}^3{He}$} abundances in a few planetary nebulae
suggest that not all low mass stars experience CBP and \hbox{$\rm{}^3{He}$}
depletion.

6. Population~II stars should experience even more severe
\hbox{$\rm{}^3{He}$} depletion than Population~I stars
($0.3 \lesssim g_3 \lesssim 0.7$), since they encounter more vigorous
cool bottom processing due to their higher hydrogen-burning temperatures.

7. This net \hbox{$\rm{}^3{He}$} destruction in stars would result in some
relaxation of the upper bound on the primordial
(D+\hbox{$\rm{}^3{He}$})/H abundance, but not by much (see, e.g.,
Olive et al.\ 1997\markcite{Oli+97};
Tosi et al.\ 1998\markcite{Tosi+98})
--- high deuterium observations in quasars are still hard to accommodate,
and only a slight relaxation is allowed on
the lower bound on the cosmic baryon density~$\Omega_b$ obtained
from Big Bang nucleosynthesis calculations.

8. For reference, we also present the effects of standard first and second
dredge-up on the helium, lithium, beryllium, and boron isotopes.  The
first dredge-up dilutions of \hbox{$\rm{}^7{Li}$}, \hbox{$\rm{}^9{Be}$},
\hbox{$\rm{}^{10}B$}, and~\hbox{$\rm{}^{11}B$}
are typically by factors of $\sim 60$, 30, 10, and 10, respectively;
there is some evidence that extra mixing on the main sequence may result
in larger dilution factors.
There is also substantial destruction of \hbox{$\rm{}^7{Li}$} by burning
during first dredge-up for solar-metallicity stars of $\lesssim 1\>M_\odot$,
and during second dredge-up for stars of $\sim 6 - 7\>M_\odot$.

9. We find that stars of $1 - 12\>M_\odot$ account for $\sim 50$\% of the
post-Big-Bang interstellar medium enrichment of~\hbox{$\rm{}^4{He}$},
with supernovae accounting for the other~50\%.

\acknowledgements

We both are indebted to Charles A. Barnes for support and encouragement as
well as insightful discussions, and Robert D. McKeown for the support
supplied by the Kellogg Radiation Laboratory.  We wish to express a special
gratitude to Charles W. Peck and Helmut A. Abt for helpful discussions and
support.  We are also grateful to G.~J.~Wasserburg for stimulating
discussions and support.  One of us (I.-J.~S.) wishes to thank Robert
F. Christy, her husband, for thoughtful comments and gentlemanly help during
the many tasks of daily life.  One of us (A.~I.~B.) wishes to thank Scott
D. Tremaine and Peter G. Martin for the support provided by the Canadian
Institute for Theoretical Astrophysics.  This work was supported in part by
a grant from the Natural Sciences and Engineering Research Council of
Canada, by a grant from the National Science Foundation
\hbox{PHY 94-20470}, by NASA grant \hbox{NAGW-3337}
to G.~J.~Wasserburg, and by a grant from the Australian Research Council.


\end{document}